\documentclass[12pt]{article}
\usepackage{epsf}

\makeatletter
\@addtoreset{equation}{section}

\renewcommand{\thefootnote}{\fnsymbol{footnote}}
\makeatother

\newcommand{\Lm}{\Lambda}
\newcommand{\lm}{\lambda}
\newcommand{\tht}{\theta}
\newcommand{\Acl}{A_{\rm cl}}
\newcommand{\Fcl}{F_{\rm cl}}
\newcommand{\am}{{\rm am}}
\newcommand{\sn}{{\rm sn}}
\newcommand{\cn}{{\rm cn}}
\newcommand{\dn}{{\rm dn}}
\newcommand{\gef}[1]{g_{{\rm eff}#1}}
\newcommand{\gefb}[1]{g_{{\rm eff}#1}^{\mbox{\scriptsize 
(B-term)}}}
\newcommand{\gefs}[1]{g_{{\rm eff}#1}^{\rm (scalar)}}

\newcommand{\hef}[1]{h_{{\rm eff}#1}}
\newcommand{\dms}{\Delta m^{2}}
\newcommand{\dMs}{\Delta M^{2}}
\newcommand{\mRsap}{m_{{\rm R},1}^{({\rm ap})2}}
\newcommand{\AR}{A_{\rm R}}
\newcommand{\AI}{A_{\rm I}}
\newcommand{\ar}[1]{a_{{\rm R},#1}}
\newcommand{\ai}[1]{a_{{\rm I},#1}}
\newcommand{\bR}[1]{b_{{\rm R},#1}}
\newcommand{\bI}[1]{b_{{\rm I},#1}}
\newcommand{\mR}[1]{m_{{\rm R},#1}}
\newcommand{\mI}[1]{m_{{\rm I},#1}}
\newcommand{\f}[2]{f^{(#1)}_{#2}}
\newcommand{\ps}[2]{\psi^{(#1)}_{#2}}
\newcommand{\CR}[1]{C_{{\rm R},#1}}
\newcommand{\Cf}[1]{C_{{\rm f},0}^{(#1)}}
\newcommand{\fm}[2]{f^{(#1)}_{{\rm m}#2}}
\newcommand{\psm}[2]{\psi^{(#1)}_{{\rm m}#2}}
\newcommand{\mm}[1]{m_{{\rm m}#1}}
\newcommand{\gm}[1]{\gamma_{(3)}^{#1}}
\newcommand{\fng}{f_{\rm NG}}
\newcommand{\psng}{\psi_{\rm NG}}
\newcommand{\amR}[2]{a_{{\rm R}#1,#2}}
\newcommand{\amI}[2]{a_{{\rm I}#1,#2}}
\newcommand{\bmR}[2]{b_{{\rm R}#1,#2}}
\newcommand{\bmI}[2]{b_{{\rm I}#1,#2}}

\newcommand{\Sat}{S_{\mbox{\scriptsize A-term}}}
\newcommand{\Ssc}{S_{\mbox{\scriptsize scalar mass}}}
\newcommand{\Sgg}{S_{\mbox{\scriptsize gaugino}}}
\newcommand{\dX}[1]{\!\!{\rm d}^{#1}X}
\newcommand{\dx}[1]{\!\!{\rm d}^{#1}x}
\newcommand{\amt}[1]{\tilde{a}_{#1}}
\newcommand{\Nm}{N}
\newcommand{\cF}{{\cal F}}
\newcommand{\cG}{{\cal G}}
\newcommand{\cH}{{\cal H}}
\newcommand{\Mf}{M}

\newcommand{\cL}{{\cal L}}
\newcommand{\del}{\partial}
\newcommand{\dy}{{\rm d}y}

\newcommand{\dls}{\partial\hspace{-6.5pt}/}
\renewcommand{\Re}{{\rm Re}}

\newcommand{\pB}{p_{\rm B}}
\newcommand{\pF}{p_{\rm F}}
\newcommand{\chF}[1]{\chi_{{\rm F}#1}}
\newcommand{\bchF}[1]{\bar{\chi}_{\rm F}^{\ #1}}
\newcommand{\bpsng}{\bar{\psi}_{\rm NG}}
\newcommand{\chng}{\chi_{\rm NG}}
\newcommand{\bchng}{\bar{\chi}_{\rm NG}}

\newcommand\be{\begin{equation}}
\newcommand\ee{\end{equation}}
\newcommand\bea{\begin{eqnarray}}
\newcommand\eea{\end{eqnarray}}

\begin{document}

\thispagestyle{empty}
%
\begin{flushright}
TIT/HEP--467 \\
UT-939\\
{\tt hep-th/0107204} \\
July, 2001 \\
\end{flushright}
\vspace{3mm}
\begin{center}
{\Large
{\bf  Simple SUSY Breaking Mechanism by Coexisting Walls
 }} 
\vskip 1.5cm


{\bf Nobuhito Maru~$^{a}$}
\footnote{\it  e-mail address: 
maru@hep-th.phys.s.u-tokyo.ac.jp},  
{\bf 
Norisuke Sakai~$^{b}$}
\footnote{\it  e-mail address: nsakai@th.phys.titech.ac.jp},
{\bf 
Yutaka Sakamura~$^{b}$}
\footnote{\it  e-mail address: sakamura@th.phys.titech.ac.jp} 
~and~~ {\bf 
Ryo Sugisaka~$^{b}$}
\footnote{\it  e-mail address: sugisaka@th.phys.titech.ac.jp}

\vskip 1.5em

{ \it   $^{a}$Department of Physics, University of Tokyo 
113-0033, JAPAN \\
and \\
$^{b}$Department of Physics, Tokyo Institute of Technology 
\\
Tokyo 152-8551, JAPAN  }
\vspace{10mm}
{\bf Abstract}\\[5mm]
{\parbox{14cm}{\hspace{5mm}
A SUSY breaking mechanism with no messenger fields is 
proposed. 
We assume that our world is on a domain wall and SUSY is broken 
only by the 
coexistence of another wall with some distance from our wall. 
We find an ${\cal N}=1$ model in four dimensions which 
admits an exact solution of 
a stable non-BPS configuration of two walls 
and 
studied its properties explicitly. 
 We work out how various soft SUSY breaking terms 
can arise in our framework. 
Phenomenological implications are briefly discussed. 
We also find that effective SUSY breaking scale 
becomes exponentially small as the distance between two 
walls grows. 
}}
\end{center}
\vfill
\newpage
\setcounter{page}{1}
\setcounter{footnote}{0}
\renewcommand{\thefootnote}{\arabic{footnote}}

\section{Introduction}\label{INTRO}
Supersymmetry (SUSY) is one of the most promising ideas to 
solve the 
 hierarchy problem in unified theories \cite{DGSW}. 
It has been noted for some years that one of the most 
important issues for 
SUSY unified theories is to understand the SUSY breaking in 
our observable 
world. 
Many models of SUSY breaking uses some kind of mediation 
of the SUSY breaking 
from the hidden sector to our observable sector. 
Supergravity provides a tree level SUSY breaking effects in 
our observable 
sector suppressed by the Planck mass $M_{\rm Pl}$ 
\cite{SugrMed}. 
Gauge mediation models uses messenger 
fields to communicate the SUSY 
breaking at the loop level in our observable sector 
\cite{DineNelson}. 

Recently there has been an active interest in  the ``Brane 
World" scenario 
where  our four-dimensional spacetime is realized on the wall 
 in higher dimensional spacetime \cite{LED,RS}. 
In order to discuss the stability of such a wall, it is often 
useful 
to consider SUSY theories as the fundamental theory. 
Moreover, SUSY theories in higher dimensions are a natural 
possibility 
in string theories. 
These SUSY theories in higher dimensions have $8$ or more 
supercharges, which should be broken partially 
if we want to have a 
phenomenologically viable SUSY unified model in four 
dimensions. 
Such a partial breaking of SUSY is nicely obtained by the 
topological 
defects \cite{WittenOlive}. 
Domain walls or other topological defects preserving 
part of the original SUSY in the fundamental theory 
are called the BPS states in SUSY theories. 
Walls have co-dimension one and typically 
 preserve half of the original SUSY, which are called 
 $1/2$ BPS states  \cite{CGR,DW,KSS}. 
 Junctions of walls have co-dimension two and typically 
preserve 
a quarter of the original SUSY \cite{AbrahamTownsend,DWJ}. 

Because of the new possibility offered by the brane world 
scenario, 
there has been a renewed interest in studies of SUSY 
breaking. 
It has been pointed out that the non-BPS topological defects 
can be a source 
of SUSY breaking \cite{DW} and an explicit realization was 
considered in the context of families localized in different 
BPS 
walls \cite{DvaliShifman}. 
Models have also been proposed with bulk fields mediating 
the SUSY breaking 
from the hidden wall to our wall on which standard model 
fields are localized 
\cite{BULK,gMSB,radion,shining}. 
The localization of the various matter wave functions in the 
extra dimensions 
was proposed to offer a natural realization of the 
gaugino-mediation of 
 the SUSY breaking \cite{KT}. 
Recently we have proposed a simple mechanism of SUSY 
breaking due to the 
coexistence of different kinds of BPS domain walls and 
proposed an efficient 
method to evaluate the SUSY breaking parameters such as 
the boson-fermion mass-splitting 
by means of overlap of wave functions involving the 
Nambu-Goldstone (NG) 
fermion \cite{MSSS}, thanks to the low-energy theorem
\cite{lee-wu,clark}. 
We have exemplified these points by taking a toy model in 
four dimensions, 
which allows an exact solution of coexisting walls with a 
three-dimensional 
effective theory. 
Although the model is only meta-stable, we were able to 
show approximate 
evaluation of the overlap allows us to determine 
the mass-splitting reliably. 

The purpose of this paper is to illustrate our idea of SUSY 
breaking due to 
the coexistence of BPS walls by taking a simple soluble 
model with a stable 
non-BPS configuration of two walls and to extend our analysis 
to more realistic case of four-dimensional effective theories. 
We also examine the consequences of our mechanism in 
detail. 

We propose a SUSY breaking mechanism which requires no 
messenger fields, 
nor  complicated SUSY breaking sector on any of the walls. 
We assume that our world is on a wall and SUSY is broken 
only by the 
coexistence of another wall with some distance from our wall. 
We find an ${\cal N}=1$ supersymmetric model in four 
dimensions 
which admits an 
exact solution of a stable non-BPS configuration of two walls 
and study its properties explicitly. 
We work out how various soft SUSY breaking terms 
can arise in our framework. 
Phenomenological implications are briefly discussed. 
We also find that effective SUSY breaking scale observed on 
our wall becomes 
exponentially small as the distance between two walls grows. 
The NG fermion is localized on the distant 
wall and its overlap with the wave functions of physical fields 
on our wall gives the boson-fermion mass-splitting of 
physical fields on our 
wall thanks to a low-energy theorem. 
We propose that this overlap provides a practical 
method to evaluate 
the mass-splitting in models with SUSY breaking due to the 
coexisting walls. 

In the next section, a model is introduced that allows a stable 
non-BPS 
two-wall configuration as a classical solution. 
We have also worked out mode expansion on the two-wall 
background, 
three-dimensional effective Lagrangian, and the single-wall 
approximation 
for the overlap of mode functions to obtain the mass-splitting. 
Matter fields are also introduced. 
Section~\ref{3D-eff} is devoted to study how various soft 
breaking terms 
arise in the three-dimensional effective theory. 
Soft breaking terms in four-dimensional effective theory are 
worked out in section~\ref{4D-eff}. 
Phenomenological implications are discussed in 
section~\ref{phenomeno-imp}. 
Additional discussion is given in section~\ref{discuss}. 
Appendix~\ref{LET-mixing} is devoted to discussing the 
low-energy theorem 
in three dimensions and the mixing matrix relating the mass 
eigenstates and superpartner states. 
Low-energy theorems in four dimensions are derived in 
Appendix~\ref{LETin4D}. 
In Appendix~\ref{V0-f1f2}, we derive a relation among the 
order parameters of 
the SUSY breaking, the energy density of the configuration 
and 
the central charge of the SUSY algebra.

\section{SUSY breaking by the coexistence of walls} 
\label{SUSY-br-coexist}
\subsection{Stable non-BPS configuration of two walls}
\label{WS:solublemodel}

\begin{figure}[h]
 \leavevmode
 \epsfysize=6cm
 \centerline{\epsfbox{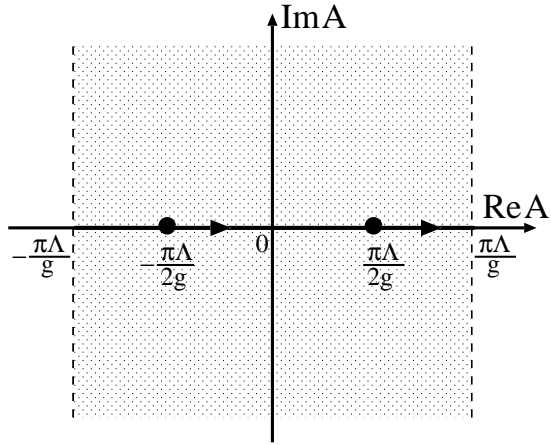}}
 \caption{The target space of the scalar field $A$. The line at 
 $\Re A=\pi\Lm/g$ and the line at $\Re A=-\pi\Lm/g$ are 
identified 
 each other.}
 \label{target-A}
\end{figure}

We will describe a simple soluble model for a stable non-BPS 
configuration that represents two-domain-wall system, 
in order to illustrate our basic ideas. 
Here we consider domain walls in four-dimensional spacetime 
to avoid inessential complications. 
We introduce a simple four-dimensional Wess-Zumino model 
as follows.\footnote{
We follow the conventions in Ref.\cite{WessBagger}
} 
\begin{eqnarray}
 &\!\!\!\cL&\!\!\!=\bar{\Phi}\Phi |_{\theta^{2}\bar{\theta}^{2}}
  +W(\Phi)|_{\theta^{2}}+{\rm h.c.},  \label{Logn} 
\qquad 
W(\Phi)
=\frac{\Lm^{3}}{g^{2}}\sin\left(\frac{g}{\Lm}\Phi\right), 
\end{eqnarray}
where $\Phi$ is a chiral superfield 
$
 \Phi(Z^\mu, \theta)
=A(Z^\mu)
+\sqrt{2}\theta\Psi(Z^\mu)
+\theta^{2}F(Z^\mu)$, 
$Z^\mu \equiv X^\mu+i\theta\sigma^\mu\bar\theta$. \\
A scale parameter $\Lm$ has a mass-dimension one and a 
coupling constant 
$g$ is dimensionless, and both of them are real positive.
In the following, we choose $y=X^{2}$ as the extra dimension 
and 
compactify it on $S^{1}$ of radius $R$.
Other coordinates are denoted as $x^{m}$ ($m=0,1,3$), {\em i.e.}, 
$X^{\mu}=(x^{m},y)$. 
The bosonic part of the model is 
\be
 \cL_{\rm 
bosonic}=-\del^{\mu}A^{\ast}\del_{\mu}A-\frac{\Lm^{4}}{g^{2}}
 \left|\cos\left(\frac{g}{\Lm}A\right)\right|^{2}.
\ee
The target space of the scalar field $A$ has a topology of a 
cylinder 
as shown in Fig.\ref{target-A}.
This model has two vacua at $A=\pm\pi\Lm/(2g)$, both lie 
on the real axis.

Let us first consider the case of the limit $R\to\infty$.
In this case, there are two kinds of BPS domain walls in this 
model.
One of them is 
\be
 \Acl^{(1)}(y)=\frac{\Lm}{g}\left\{2\tan^{-1}e^{\Lm(y-y_{1})}
 -\frac{\pi}{2}
 \right\},
 \label{eq:first_wall}
\ee
which interpolates the vacuum at $A=-\pi\Lm/(2g)$ to that 
at 
$A=\pi\Lm/(2g)$ as $y$ increases from $y=-\infty$ to 
$y=\infty$.
The other wall is 
\be
 \Acl^{(2)}(y)=\frac{\Lm}{g}\left\{-2\tan^{-1}e^{-\Lm(y-y_{2})}+
\frac{3\pi}{2}
 \right\},
 \label{eq:second_wall}
\ee
which interpolates the vacuum at $A=\pi\Lm/(2g)$ to that at 
$A=3\pi\Lm/(2g)=-\pi\Lm/(2g)$.
Here $y_{1}$ and $y_{2}$ are integration constants and 
represent 
the location of the walls along the extra dimension. 
The four-dimensional supercharge $Q_{\alpha}$ can be 
decomposed into 
two two-component Majorana supercharges $Q^{(1)}_{\alpha}$ 
and 
$Q^{(2)}_{\alpha}$ which can be regarded as supercharges in 
three dimensions 
\be
 Q_{\alpha}=\frac{1}{\sqrt{2}}(Q^{(1)}_{\alpha}+iQ^{(2)}_{\alpha}).
\ee
Each wall breaks a half of the bulk supersymmetry: 
 $Q^{(1)}_{\alpha}$ is broken by $\Acl^{(2)}(y)$, and $Q^{(2)}_{\alpha}$ 
 by $\Acl^{(1)}(y)$.
Thus all of the bulk supersymmetry will be broken if these 
walls coexist.

We will consider such a two-wall system to study the SUSY 
breaking 
effects in the low-energy three-dimensional theory on the 
background. 
The field configuration of the two walls will wrap around the 
cylinder in 
the target space of $A$ as $y$ increases from $0$ to $2\pi 
R$. 
Such a configuration should be a solution of the equation of 
motion, 
\be
 \del^{\mu}\del_{\mu}A+\frac{\Lm^{3}}{g}\sin\left(\frac{g}{\Lm}
A^{\ast}\right)
 \cos\left(\frac{g}{\Lm}A\right)=0. \label{EOM1}
\ee
We can easily show that 
the minimum energy static configuration with unit winding 
number 
 should be real. 
We find that a general real static solution of Eq.(\ref{EOM1}) 
that depends 
only on $y$ is 
\be
 \Acl(y)=\frac{\Lm}{g}\am\left(\frac{\Lm}{k}(y-y_{0}),k\right), 
 \label{A_classical}
\ee
where $k$ and $y_{0}$ are real parameters and the function 
$\am(u,k)$ 
denotes the amplitude function, which is defined 
as an inverse function of 
\be
 u(\varphi)=\int_{0}^{\varphi}\frac{{\rm d}\theta}
 {\sqrt{1-k^{2}\sin^{2}\theta}}.
\ee
If $k>1$, it becomes a periodic function with the period 
$4K(1/k)/\Lm$, where the function $K(k)$ is the complete 
elliptic integral 
of the first kind.
If $k<1$, the solution $\Acl(y)$ is a monotonically increasing 
function with 
\be
 \Acl\left(y+{4kK(k) \over \Lambda}\right)= \Acl(y)+ 2 
\pi{\Lambda \over g}.
\ee
\begin{figure}[t]
 \leavevmode
 \epsfysize=8cm
 \centerline{\epsfbox{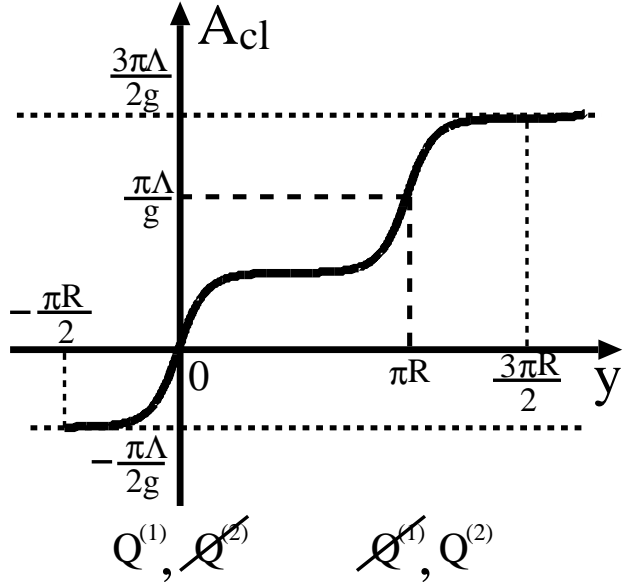}}
 \caption{The profile of the classical solution $\Acl(y)$. 
 The dotted lines $A=-\pi\Lm/(2g)$ and $A=3\pi\Lm/(2g)$ 
are identified.
 }
 \label{profile-Acl}
\end{figure}
This is the solution that we want.
Since the field $A$ is an angular variable 
$A=A+2\pi\Lambda/g$, 
we can choose the compactified radius 
$2\pi R=4kK(k)/\Lm$ so that the classical field configuration 
$\Acl(y)$ contains two walls and becomes 
periodic modulo $2\pi\Lambda/g$. 
We shall take $y_{0}=0$ to locate one of the walls at $y=0$. 
Then we find that the other wall is located at the anti-podal 
point 
 $y=\pi R$ of the compactified circle. 
We have computed the energy of a superposition of the first 
wall 
$\Acl^{(1)}(y)$ located at $y=y_1$in Eq.(\ref{eq:first_wall}) 
 and the second wall $ \Acl^{(2)}(y)$ located at $y=y_2$ 
in Eq.(\ref{eq:second_wall}). 
This energy can be regarded as a potential between two walls 
in the adiabatic 
approximation and has a peak at $|y_1-y_2|=0$ implying that 
two walls 
experience a repulsion. 
This is in contrast to a BPS configuration of two walls which 
should exert 
no force between them. 
Thus we can explain that the second wall is settled at the 
anti-podal point $y=\pi R$ in our stable non-BPS 
configuration because of 
the repulsive force between two walls. 

In the limit of $R\to\infty$, {\em i.e.}, $k\to 1$, $\Acl(y)$ 
approaches to 
the BPS configuration $\Acl^{(1)}(y)$ with $y_1=0$ near $y=0$, which 
preserves $Q^{(1)}$, 
and to $\Acl^{(2)}(y)$ with $y_2=\pi R$ near $y=\pi R$, which preserves 
$Q^{(2)}$. 
The profile of the classical solution $\Acl(y)$ is shown 
in Fig.\ref{profile-Acl}. 
We will refer to the wall at $y=0$ as ``our wall'' and the wall at 
$y=\pi R$ as ``the other wall''.

\subsection{The fluctuation mode expansion}
Let us consider the fluctuation fields around the background 
$\Acl(y)$, 
\bea
 A(X)&\!\!=&\!\!\Acl(y)+\frac{1}{\sqrt{2}}(\AR(X)+i\AI(X)), 
 \nonumber\\
 \Psi_{\alpha}(X)&\!\!=&\!\!\frac{1}{\sqrt{2}}(\Psi_{\alpha}^{(1)}(
X)
 +i\Psi_{\alpha}^{(2)}(X)). \label{fluc_fields}
\eea
To expand them in modes, we define 
the mode functions as solutions of equations:
\bea
 \left\{-\del_{y}^{2}-\Lm^{2}\cos\left(\frac{2g}{\Lm}\Acl(y) 
 \right)\right\}
 \bR{p}(y)&\!\!=&\!\!\mR{p}^{2}\bR{p}(y), \nonumber\\
 \{-\del_{y}^{2}+\Lm^{2}\}\bI{p}(y)&\!\!=&\!\!\mI{p}^{2}(y)\bI{p}(y
), 
 \label{boson_mode_eq}
\eea
\bea
 \left\{-\del_{y}-\Lm\sin\left(\frac{g}{\Lm}\Acl(y)\right)\right
\}
 \f{1}{p}(y)&\!\!=&\!\!m_{p}\f{2}{p}(y), \nonumber\\
 \left\{\del_{y}-\Lm\sin\left(\frac{g}{\Lm}\Acl(y) 
 \right) \right\}
 \f{2}{p}(y)&\!\!=&\!\!m_{p}\f{1}{p}(y). \label{fermion_mode_eq}
\eea
The four-dimensional fluctuation fields can be expanded as 
\be
 \AR(X)=\sum_{p}\bR{p}(y)\ar{p}(x),\;\;\;
 \AI(X)=\sum_{p}\bI{p}(y)\ai{p}(x), 
\label{eq:boson_mode_decomp}
\ee
\be
 \Psi^{(1)}(X)=\sum_{p}\f{1}{p}(y)\ps{1}{p}(x),\;\;\;
 \Psi^{(2)}(X)=\sum_{p}\f{2}{p}(y)\ps{2}{p}(x). 
 \label{eq:fermion_mode_decomp}
\ee
As a consequence of the linearized equation of motion, the 
coefficient 
$\ar{p}(x)$ and $\ai{p}(x)$ are scalar fields in 
three-dimensional effective theory with masses $\mR{p}$ and 
$\mI{p}$,
and $\ps{1}{p}(x)$ and $\ps{2}{p}(x)$ are three-dimensional 
spinor fields with masses $m_{p}$, respectively.

\begin{figure}[t]
 \leavevmode
 \epsfysize=6cm
 \centerline{\epsfbox{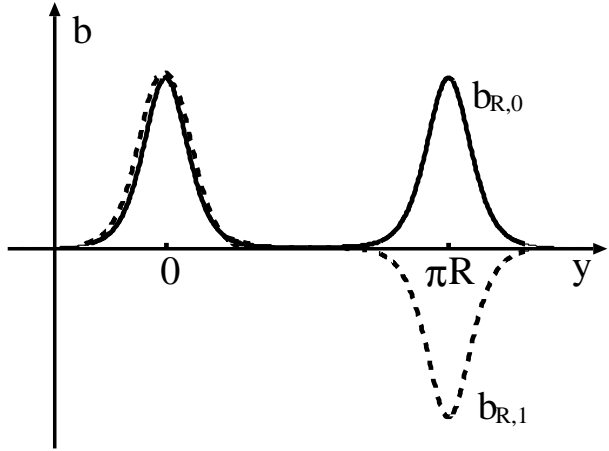}}
 \caption{The mode functions for the bosonic modes $\ar{0}$ 
 and $\ar{1}$. The solid line represents the profile of 
$\bR{0}(y)$ and 
 the dashed line is that of $\bR{1}(y)$.
 }
 \label{boson-mode}
\end{figure}
Exact mode functions and mass-eigenvalues are known for 
several light modes 
of $\bR{p}(y)$, 
\bea
 \bR{0}(y)&\!\!=&\!\!\CR{0}\dn\left(\frac{\Lm y}{k},k\right), 
 \;\;\; \mR{0}^{2}=0, \nonumber \\
 \bR{1}(y)&\!\!=&\!\!\CR{1}\cn\left(\frac{\Lm y}{k},k\right), \;
\;\; 
 \mR{1}^{2}=\frac{1-k^{2}}{k^{2}}\Lm^{2}, \nonumber \\
 \bR{2}(y)&\!\!=&\!\!\CR{2}\sn\left(\frac{\Lm y}{k},k\right), \;
\;\; 
 \mR{2}^{2}=\frac{\Lm^{2}}{k^{2}}, \label{bosonR_mode_fnc}
\eea
where functions $\dn(u,k)$, $\cn(u,k)$, $\sn(u,k)$ are the 
Jacobi's 
elliptic functions and $\CR{p}$ are normalization factors. 
{}For $\bI{p}(y)$, we can find all the eigenmodes 
\be
 \bI{p}(y)=\frac{1}{\sqrt{2\pi R}}{\rm e}^{i{p \over R}y}
,\;\;\; 
 \mI{p}^{2}=\Lm^{2}+\frac{p^{2}}{R^{2}},\;\;\;
 ( p \in {\bf Z} 
 ).
\ee
The massless field $\ar{0}(x)$ is the Nambu-Goldstone (NG) 
boson 
for the breaking of the translational invariance in the extra 
dimension.
The first massive field $\ar{1}(x)$ corresponds to the 
oscillation of the 
background wall around the anti-podal equilibrium point and 
hence becomes 
massless in the limit of $R\rightarrow \infty$. 
All the other bosonic fields remain massive in that limit. 

{}For fermions, only zero modes are known explicitly,
\be
 \f{1}{0}(y)=C_{0}\left\{\dn\left(\frac{\Lm y}{k},k\right)
 +k\cn\left(\frac{\Lm y}{k},k\right)\right\}, \;\;\;
 \f{2}{0}(y)=C_{0}\left\{\dn\left(\frac{\Lm y}{k},k\right)
 -k\cn\left(\frac{\Lm y}{k},k\right)\right\}, 
\label{fermion_mode_fnc}
\ee
where $C_{0}$ is a normalization factor.
These fermionic zero modes are the NG fermions for the 
breaking of 
$Q^{(1)}$-SUSY and $Q^{(2)}$-SUSY, respectively. 

\begin{figure}[t]
 \epsfysize=5cm
 \centerline{\epsfbox{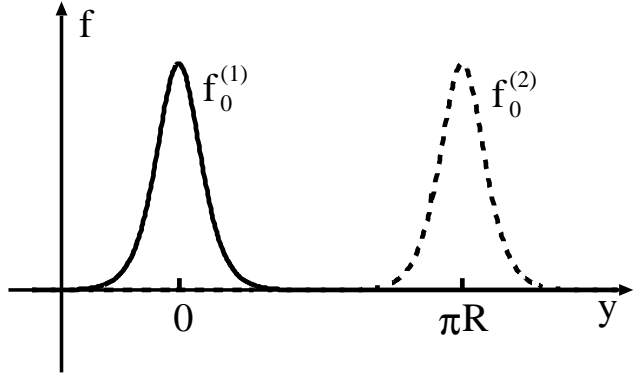}}
 \caption{The mode functions for fermionic zero-modes 
$\ps{1}{0}$ and
 $\ps{2}{0}$. The solid line represents the profile of $\f{1}{0}(y)
$ and 
 the dashed line is that of $\f{2}{0}(y)$.
 }
 \label{fermion-mode}
\end{figure}

Thus there are four fields which are massless or become 
massless in the limit 
of $R\rightarrow \infty$: 
$\ar{0}(x)$, $\ar{1}(x)$, $\ps{1}{0}(x)$ and $\ps{2}{0}(x)$.
The profiles of their mode functions are shown in 
Fig.\ref{boson-mode} and 
Fig.\ref{fermion-mode}.
Other fields are heavier and have masses of the order of 
$\Lm$. 

In the following discussion, we will concentrate ourselves on 
the breaking of the $Q^{(1)}$-SUSY, which is approximately 
preserved 
by our wall at $y=0$.
So we call the field $\ps{2}{0}(x)$ the NG fermion 
in the rest of the paper.

\subsection{Three-dimensional effective Lagrangian}
We can obtain a three-dimensional effective Lagrangian 
by substituting the mode-expanded fields 
Eq.(\ref{eq:boson_mode_decomp}) 
and Eq.(\ref{eq:fermion_mode_decomp}) into the Lagrangian 
(\ref{Logn}), 
and carrying out an integration over $y$ 
\begin{eqnarray}
 \cL^{(3)}&\!\!\!=&\!\!\!
 -V_{0}-\frac{1}{2}\del^{m}\ar{0}\del_{m}\ar{0}
 -\frac{1}{2}\del^{m}\ar{1}\del_{m}\ar{1}-\frac{i}{2}\ps{1}{0}
 \dls\ps{1}{0}
 -\frac{i}{2}\ps{2}{0}\dls\ps{2}{0} \nonumber \\
 &\!\!\!&\!\!\!
 -\frac{1}{2}\mR{1}^{2}\ar{1}^{2}+\gef{}\ar{1}\ps{1}{0}\ps{2}{0}
 +\cdots, 
 \label{effthry}
\end{eqnarray}
where $\dls\equiv\gamma^{m}_{(3)}\del_{m}$ and 
an abbreviation denotes terms involving heavier fields and 
higher-dimensional terms. 
Here 
 $\gamma$-matrices in three dimensions are defined by 
 $\left(\gamma^{m}_{(3)}\right)
 \equiv\left(-\sigma^2, i\sigma^3, -i\sigma^1\right)$. 
The vacuum energy $V_0$ is given by the energy density of 
the background 
and thus 
\begin{equation}
 V_{0}\equiv \int^{\pi R}_{-\pi R}{\rm 
d}y\left\{\left(\del_{y}\Acl\right)
 +\frac{\Lm^{4}}{g^{2}}\cos^{2}\left(\frac{g}{\Lm}\Acl 
 \right) \right\}
 =\frac{\Lm^{3}}{g^{2}k}\int_{-2K(k)}^{2K(k)}{\rm d}u\left\{
 (1+k^{2})-2k^{2}\sn^{2}(u,k)\right\},
\label{eq:vacuum_energy}
\end{equation}
and the effective Yukawa coupling $\gef{}$ is 
\begin{equation}
 \gef{}\equiv\frac{g}{\sqrt{2}}\int^{\pi R}_{-\pi R}{\rm d}y \, 
 \cos\left(\frac{g}{\Lm}\Acl(y)\right) \bR{1}(y)\f{1}{0}(y)\f{2}{0}
(y)
 =\frac{g}{\sqrt{2}}\frac{C_{0}^{2}}{\CR{1}}(1-k^{2}).
 \label{geff}
\end{equation}

In the limit of $R\to\infty$, the parameters $\mR{1}$ and 
$\gef{}$ vanish 
and thus we can redefine the bosonic massless fields 
as 
\be
 \left(\begin{array}{c}a^{(1)}_{0} \\ a^{(2)}_{0}\end{array}\right)
 =\frac{1}{\sqrt{2}}\left(\begin{array}{cc} 1 & 1 \\ -1 & 1 
\end{array}\right)
 \left(\begin{array}{c}\ar{0} \\ \ar{1}\end{array}\right). 
 \label{boson-mixing}
\ee
In this case, the fields $a^{(1)}_{0}(x)$ and $\ps{1}{0}(x)$ 
form a supermultiplet 
for $Q^{(1)}$-SUSY and their mode functions are both 
localized on our wall. 
The fields $a^{(2)}_{0}(x)$ and $\ps{2}{0}(x)$ are singlets 
for $Q^{(1)}$-SUSY and are localized on the other 
wall.\footnote{
The modes $a^{(2)}_{0}(x)$ and $\ps{2}{0}(x)$ form a 
supermultiplet for 
$Q^{(2)}$-SUSY.
} 

When the distance between the walls $\pi R$ is finite, 
$Q^{(1)}$-SUSY is broken and the mass-splittings between 
bosonic and fermionic 
modes are induced.
The mass squared $\mR{1}^{2}$ in Eq.(\ref{effthry}) 
corresponds to 
the difference of the mass squared $\dms$ 
between $a^{(1)}_{0}(x)$ and $\ps{1}{0}(x)$ 
since the fermionic mode $\ps{1}{0}(x)$ is massless.
Besides the mass terms, we can read off the SUSY breaking 
effects 
from the Yukawa couplings like $\gef{}$.

We have noticed in Ref.\cite{MSSS} that these two SUSY 
breaking parameters, 
$\mR{1}$ and $\gef{}$, 
are related by the low-energy theorem associated with the 
spontaneous breaking of SUSY. 
In our case, the low-energy theorem becomes 
\begin{equation}
 \frac{\gef{}}{\mR{1}^{2}}=\frac{1}{2f}. \label{GTR-gef}
\end{equation}
where $f$ is an order parameter of the SUSY breaking, and it 
is given 
by the square root of the vacuum (classical background) 
energy density $V_0$ in Eq.(\ref{eq:vacuum_energy}). 
The low-energy theorem in three dimensions is briefly 
explained 
in Appendix~\ref{LET-in-3D}.
Since the superpartner of the fermionic field $\ps{1}{0}(x)$ 
is a mixture of mass-eigenstates, 
we had to take into account the mixing 
Eq.(\ref{boson-mixing}). 
The mixing in general situation is discussed and is applied to 
the present case in Appendix~\ref{sp-me} and 
\ref{application}.

\begin{figure}[t]
 \leavevmode
 \epsfxsize=9cm
 \epsfysize=6cm
 \centerline{\epsfbox{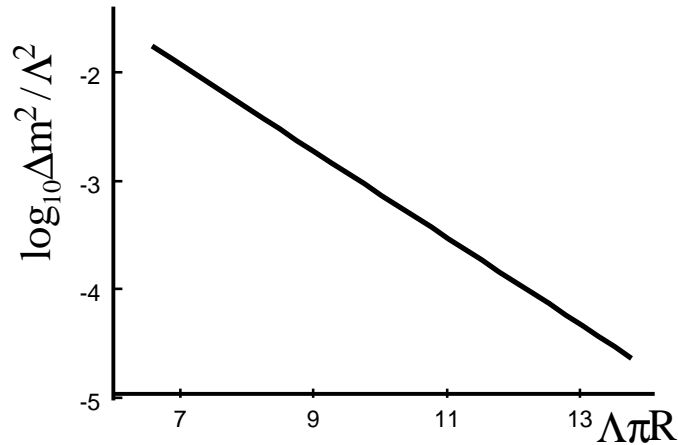}}
 \caption{The logarithm of the mass-splitting as a function 
 of the distance between the wall.
 The horizontal axis is the wall distance normalized by 
$1/\Lm$.
 }
\label{geff-R}
\end{figure}

Fig.\ref{geff-R} shows the mass-splitting $\dms$ 
as a function of the wall distance $\pi R$.
As this figure shows, the mass-splitting decays exponentially 
as the wall distance increases.
This is one of the characteristic features of our SUSY 
breaking mechanism. 
This fact can be easily understood by remembering the profile 
of each modes. 
Note that the mass-splitting $\dms(=\mR{1}^{2})$ is 
proportional 
to the effective Yukawa coupling constant $\gef{}$, 
which is represented by an overlap integral of the mode 
functions. 
Here the mode functions of the fermionic field $\ps{1}{0}(x)$ 
and 
its superpartner are both localized 
on our wall, and that of the NG fermion $\ps{2}{0}(x)$ is 
localized 
on the other wall.
Therefore the mass-splitting becomes exponentially small 
when the distance 
between the walls increases, because of exponentially 
dumping tails 
of the mode functions.

\subsection{Single-wall approximation} \label{single-wall_ap}
Next we will propose a practical method of estimation for the 
mass-splittings.
We often encounter the case where single-BPS-domain-wall 
solutions 
are known but exact two-wall configurations are not.
This is because the latter are solutions of a second order 
differential 
equation, namely the equation of motion, while the former are 
solutions of first order differential equations, namely BPS 
equations.
We can estimate the mass-splitting
by using only informations on the single-wall background, 
even if two-wall configurations are not known.
As mentioned in the previous subsection, the mass-splitting 
$\dms$ 
is related to the coupling constant $\gef{}$ and 
the order parameter $f$.
So we can estimate $\dms$ by calculating $\gef{}$ and $f$.
 
When two walls are far apart, the energy of 
the background $V_{0}$ in Eq.(\ref{eq:vacuum_energy}) 
can be well-approximated by the sum of 
those of our wall and of the other wall.
\be
 V_{0}\simeq 2\frac{\Lm^{3}}{g^{2}}\int_{-\infty}^{\infty}{\rm d}u
 \left\{2-2\tanh^{2}u\right\}
 =\frac{8\Lm^{3}}{g^{2}}. \label{V0-ap}
\ee

Considering the profiles of background and mode functions, 
we can see that the main contributions to the overlap 
integral of $\gef{}$ 
come from neighborhood of our wall and the other wall. 
These two regions give the same numerical contributions to 
the integral, 
including their signs.
Thus we can obtain $\gef{}$ by calculating the overlap 
integral of 
approximate background and mode functions 
which well approximate their behaviors near our wall, 
and multiplying it by two.

In the neighborhood of our wall, the two-wall background 
$\Acl(y)$ 
can be well approximated by the single-wall background 
$\Acl^{(1)}(y)$ with $y_1=0$. 
So,  
\be
 \cos\left(\frac{g}{\Lm}\Acl(y)\right)\simeq
 \cos\left(\frac{g}{\Lm}\Acl^{(1)}(y)\right)
 =\frac{1}{\cosh(\Lm y)}. \label{Acl-ap}
\ee

Next, we will proceed to the approximation of mode functions.
From the mode equations in Eq.(\ref{fermion_mode_eq}), we 
can express 
the zero-modes $\f{1}{0}(y)$ and $\f{2}{0}(y)$ as 
\bea
 \f{1}{0}(y)&\!\!=&\!\!\Cf{1}e^{-\int_{0}^{y}\dy'
 \Lm\sin\left(\frac{g}{\Lm}\Acl(y')\right)}, \label{f10} \\
 \f{2}{0}(y)&\!\!=&\!\!\Cf{2}e^{\int_{0}^{y}\dy'
 \Lm\sin\left(\frac{g}{\Lm}\Acl(y')\right)},  \label{f20}
\eea
where $\Cf{1}$ and $\Cf{2}$ are normalization factors.

Since the function $\f{1}{0}(y)$ has its support mainly on our 
wall, 
it is simply approximated near our wall by 
\be
 \f{1}{0}(y)\simeq\Cf{1}e^{-\int_{0}^{y}\dy'
 \Lm\sin\left(\frac{g}{\Lm}\Acl^{(1)}(y')\right)}
 =\frac{\Cf{1}}{\cosh(\Lm y)}. \label{f10ap}
\ee
Then we can determine $\Cf{1}=\sqrt{\Lm/2}$ 
by the normalization condition.

Similarly, the mode $\f{2}{0}(y)$ can be approximated near our 
wall by 
\be
 \f{2}{0}(y)\simeq\Cf{2}e^{\int_{0}^{y}\dy'
 \Lm\sin\left(\frac{g}{\Lm}\Acl^{(1)}(y')\right)}
 =\Cf{2}\cosh(\Lm y). \label{f20ap1}
\ee
Unlike the case of $\f{1}{0}(y)$, however, we cannot 
determine $\Cf{2}$ 
by using this approximate expression because the mode 
$\f{2}{0}(y)$ 
is localized mainly on the other wall. 
Here it should be noted that $\f{2}{0}(y)=\f{1}{0}(y-\pi R)$ 
from Eq.(\ref{fermion_mode_eq}) and the property of the 
background: 
$\Acl(y-\pi R)=\Acl(y)-\pi\Lm/g$.
Thus, 
\bea
 \f{2}{0}(y)&\!\!=&\!\!\Cf{1}e^{-\int_{0}^{y-\pi R}\dy'
 \Lm\sin\left(\frac{g}{\Lm}\Acl(y')\right)}
 =\Cf{1}e^{\int_{\pi R}^{y}\dy'
 \Lm\sin\left(\frac{g}{\Lm}\Acl(y')\right)} \nonumber\\
 &\!\!=&\!\!\Cf{1}e^{-\int_{0}^{\pi R}\dy'
 \Lm\sin\left(\frac{g}{\Lm}\Acl(y')\right)}
 e^{\int_{0}^{y}\dy'\Lm\sin\left(\frac{g}{\Lm}\Acl(y')\right)},
\eea
and we can obtain a relation:
\be
 \Cf{2}=\Cf{1}e^{-\int_{0}^{\pi R}\dy'\Lm\sin\left(\frac{g}{\Lm}
\Acl(y')
 \right)}.
\ee
In the region of $y\in [0,\pi R]$, the background is well 
approximated by 
\be
 \Acl(y)\simeq\left\{\begin{array}{l}
  \Acl^{(1)}(y) \;\;\;(0\leq y < \frac{\pi R}{2}) \\
  \Acl^{(2)}(y) \;\;\;(\frac{\pi R}{2} < y \leq \pi R) 
\end{array}\right.
\ee
with $y_1=0$ and $y_2=\pi R$, and thus 
\bea
 \sin\left(\frac{g}{\Lm}\Acl(y)\right)&\!\!\simeq&\!\!
 \left\{\begin{array}{l}
  \sin\left(\frac{g}{\Lm}\Acl^{(1)}(y)\right)=\tanh(\Lm y) \;\;\;
  (0\leq y < \frac{\pi R}{2}) \\
  \sin\left(\frac{g}{\Lm}\Acl^{(2)}(y)\right)=-\tanh(\Lm(y-\pi 
R)) \;\;\;
  (\frac{\pi R}{2} < y \leq \pi R) \end{array}\right. 
\nonumber\\
 &\!\!\simeq&\!\! \tanh(\Lm y)-\tanh(\Lm(y-\pi R))-1.
\eea
Thus the normalization factor can be estimated as
\be
 \Cf{2}=\Cf{1}\frac{e^{\Lm\pi R}}{\cosh^{2}\Lm\pi R}
 \simeq 2\sqrt{2\Lm}e^{-\Lm\pi R}.
\ee
Here we used the fact that $\Cf{1}=\sqrt{\Lm/2}$ and 
$\Lm\pi R\gg 1$.
As a result, the mode function of the NG fermion $\f{2}{0}(y)$ 
can be 
approximated near our wall by
\be
 \f{2}{0}(y)=2\sqrt{2\Lm}e^{-\Lm\pi R}\cosh(\Lm y). 
\label{f20ap2}
\ee

In the limit of $R\to\infty$, the $Q^{(1)}$-SUSY is recovered 
and thus 
the mode function of the bosonic field $a^{(1)}_0(x)$ 
in Eq.(\ref{boson-mixing}), 
$b^{(1)}_0(y)$, is identical to $\f{1}{2}(y)$. 
However, when the other wall exist at finite distance from our 
wall, 
this bosonic field is mixed with the field $a^{(2)}_0(x)$ 
localized on the other wall.
Because the masses of these two fields $a^{(1)}_0(x)$ and 
$a^{(2)}_0(x)$ are 
degenerate (both are massless), the maximal mixing occurs.
(See Eq.(\ref{boson-mixing}).)
\be
 \left(\begin{array}{c} \bR{0} \\ \bR{1} \end{array}\right)
 =\frac{1}{\sqrt{2}}\left(\begin{array}{cc} 1 & -1 \\ 1 & 1 
\end{array}
 \right)\left(\begin{array}{c} b^{(1)}_0 \\ b^{(2)}_0 
\end{array}\right),
\ee
where $b^{(2)}_0(y)$ is the mode function of $a^{(2)}_0(y)$.
Thus the mode function of the mass-eigenmode $\bR{1}(y)$ 
is 
approximated near our wall by 
\be
 \bR{1}(y)\simeq\frac{1}{\sqrt{2}}\f{1}{0}(y)
 \simeq \frac{\sqrt{\Lm}}{2}\frac{1}{\cosh(\Lm y)}. 
\label{br1-ap}
\ee
Then by using Eqs.(\ref{Acl-ap}), (\ref{f10ap}), 
(\ref{f20ap2}) and (\ref{br1-ap}), 
we can obtain the effective Yukawa coupling constant $\gef{}
$,
\be
 \gef{}\simeq 2g\sqrt{2\Lm}e^{-\Lm\pi R}.
\ee
As a result, the approximate mass-splitting value $\mRsap$ 
is estimated as
\be
 \mRsap=2f\gef{}=16\Lm^{2}e^{-\Lm\pi R},
\ee
by using Eq.(\ref{V0-ap}) and the low-energy theorem 
Eq.(\ref{GTR-gef}).
From this expression, we can explicitly see its exponential 
dependence of 
the distance between the walls.
We call this method of estimation the single-wall 
approximation.

\begin{figure}[t]
\leavevmode
\epsfysize=6cm
\centerline{\epsfbox{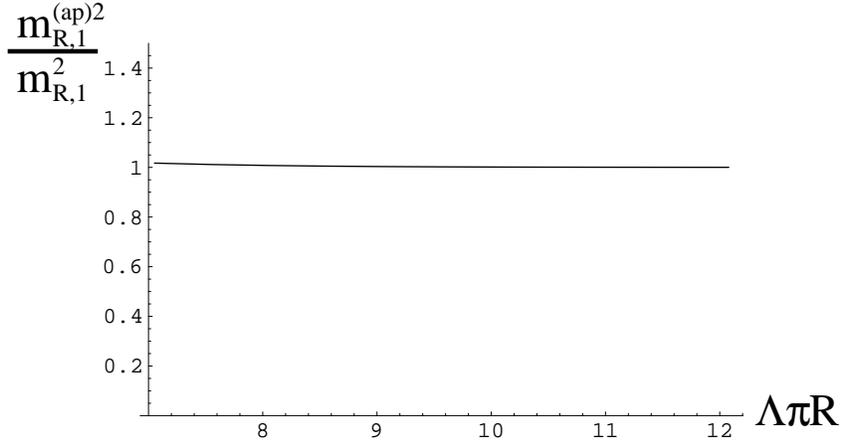}}
\caption{The ratio of the approximate value $\mRsap$ to the 
exact one 
$\mR{1}^{2}$ as a function of the wall distance $\pi R$.
The horizontal axis is normalized by $1/\Lm$.
}
\label{rto-mRs}
\end{figure}

In our model, we know the exact mass-eigenvalue $\mR{1}^{2}
$.
So we can check the validity of the above approximation by 
comparing 
the approximate value $\mRsap$ and the exact one 
$\mR{1}^{2}$.
Fig.\ref{rto-mRs} shows the ratio of $\mRsap$ to $\mR{1}^{2}
$ 
as a function of the wall distance $\pi R$.
As this figure shows, we can conclude that the single-wall 
approximation 
is very well.

\subsection{Matter fields} \label{matter_fields}
Let us introduce a matter chiral superfield 
\begin{equation}
 \Phi_{\rm m} =A_{\rm m}+\sqrt{2}\theta\Psi_{\rm m} + 
\theta^{2}F_{\rm m},
\end{equation} 
interacting with $\Phi$ 
in the original Lagrangian (\ref{Logn}) through an additional 
superpotential 
\begin{equation}
W_{\rm 
int}=-h\frac{\Lm}{g}\sin\left(\frac{g}{\Lm}\Phi\right)\Phi_{\rm 
m}^{2}
=-h\Phi\Phi_{\rm m}^{2}+\cdots, \label{Wint}
\end{equation} 
which will be treated as a small perturbation\footnote{
We can take the interaction like $W_{\rm int}=-h\Phi\Phi_{\rm 
m}^{2}$ 
as in Ref.\cite{MSSS} in order to localize the mode function 
of the 
light matter fields on our wall. 
The choice of $W_{\rm int}$ like Eq.(\ref{Wint}) is completely 
a matter of convenience.
}. 

Let us decompose the matter fermion $\Psi_{\rm m}(X)$ into 
two real two-component 
spinors $\Psi_{{\rm m}\alpha}^{(1)}(X)$ and 
$\Psi_{{\rm m}\alpha}^{(2)}(X)$ as 
$ \Psi_{{\rm m}\alpha}=
 (\Psi_{{\rm m}\alpha}^{(1)}+i\Psi_{{\rm m}\alpha}^{(2)})/\sqrt{2}
$. 
Then these fluctuation fields can be expanded by the mode 
functions 
as follows.
\be
 \Psi_{\rm m}^{(1)}(X)=\sum_{p}\fm{1}{p}(y)\psm{1}{p}(x),\;\;\;
 \Psi_{\rm m}^{(2)}(X)=\sum_{p}\fm{2}{p}(y)\psm{2}{p}(x).
\ee
The mode equations are defined as 
\bea
 \left\{-\del_{y}-\frac{2h}{g}\Lm\sin\left(\frac{g}{\Lm}\Acl 
 \right) \right\}
 \fm{1}{p}(y)&\!\!=&\!\!\mm{p}\fm{2}{p}(y), \nonumber\\
 \left\{\del_{y}-\frac{2h}{g}\Lm\sin\left(\frac{g}{\Lm}\Acl 
 \right) \right\}
 \fm{2}{p}(y)&\!\!=&\!\!\mm{p}\fm{1}{p}(y).
\eea
Thus zero-modes on the two-wall background 
(\ref{A_classical}) 
can be solved exactly 
\bea
 \fm{1}{0}(y)&\!\!=&\!\!C_{{\rm m}0}\left\{\dn\left(\frac{\Lm y}
{k},k\right)
 +k\cn\left(\frac{\Lm y}{k},k\right)\right\}^{\frac{2h}{g}}, 
\nonumber\\
 \fm{2}{0}(y)&\!\!=&\!\!C_{{\rm m}0}\left\{\dn\left(\frac{\Lm y}
{k},k\right)
 -k\cn\left(\frac{\Lm y}{k},k\right)\right\}^{\frac{2h}{g}}.
\eea
the mode $\fm{1}{0}(y)$ is localized on our wall and 
the mode $\fm{2}{0}(y)$ is on the other wall. 

Besides these zero-modes, there are 
several light modes of $\Phi_{\rm m}$ localized on our wall 
when the coupling $h$ is taken to be larger than $g$. 
Those non-zero-modes can be obtained analytically in the 
limit of 
$R\to\infty$. 
For example, the low-lying mass-eigenvalues are discrete at 
$\mm{p}^{2}=p(-p+4h/g)\Lm^{2}$ with $p=0,1,2,\cdots <2h/g$, 
and 
the corresponding mode functions $\fm{1}{p}(y)$ for the fields 
$\psm{1}{p}(x)$ are
\be
 \fm{1}{p}(y)=\frac{C_{{\rm m}p}}{[\cosh(\Lm y)]^{\frac{2h}{g}-p}}
 F\left(-p,1-p+\frac{4h}{g},1-p+\frac{2h}{g};\frac{1-\tanh(\Lm 
y)}{2}\right),
 \label{hypergeometric}
\ee
where $F(\alpha,\beta,\gamma;z)$ is the hypergeometric 
function and 
$C_{{\rm m}p}$ is normalization factors.
The mode functions $\fm{2}{p}(y)$ for the fields $\psm{2}{p}(x)
$ 
have forms similar to those of $\fm{1}{p}(y)$.

Although we do not know the exact mass-eigenvalues and 
mode functions 
in the case that the wall distance is finite, 
we can estimate the boson-fermion mass-splittings $\dms_{p}
$ by using 
the single-wall approximation discussed in the previous 
subsection.
{}For example, let us estimate the mass-splitting between 
$\psm{1}{p}(x)$ and its superpartner $a_{{\rm m}p}^{(1)}(x)$.
After including an interaction like Eq.(\ref{Wint}), 
the effective Lagrangian has the following Yukawa coupling 
terms.
\be
 \cL_{\rm int}^{(3)}=\sum_{p}\hef{p}a_{{\rm 
m}p}^{(1)}\psm{1}{p}\ps{2}{0}
 +{\rm h.c.}+\cdots, \label{matter-Yukawa}
\ee
\be
 \hef{p}=\sqrt{2}h\int_{-\pi R}^{\pi R}\dy
 \cos\left(\frac{g}{\Lm}\Acl(y)\right)
 b_{{\rm m}p}^{(1)}(y)\fm{1}{p}(y)\f{2}{0}(y). \label{hefpq}
\ee
Just like the case of $a^{(1)}_0(x)$ and $a^{(2)}_0(x)$, the 
degenerate 
states $a_{{\rm m}p}^{(1)}(x)$ and $a_{{\rm m}p}^{(2)}(x)$ are 
maximally 
mixed with each other and their masseigenvalues split into 
two different 
values $m_{{\rm mR}2p}$ and $m_{{\rm mR}(2p+1)}$. 
By calculating the effective coupling $\hef{p}$ in 
Eq.(\ref{hefpq}) 
in the single-wall approximation, 
we can obtain the following mass-splitting. (See 
Appendix~\ref{E-SWA}.)
\be
 \dms_p\equiv\frac{m_{{\rm mR}2p}^2+m_{{\rm mR}(2p+1)}^2}{2}
-\mm{p}^2.
\ee

Thanks to the approximate supersymmetry, $Q^{(1)}$-SUSY, 
we can use 
the mode function in Eq.(\ref{hypergeometric}) as both 
$\fm{1}{p}(y)$ and 
$b_{{\rm m}p}^{(1)}(y)$. 
Then we obtain the mass-splitting in the single wall 
approximation 
\be
 \dms_{p}=\sqrt{2}f\hef{p}=\frac{16h}{g}\Lm^{2}e^{-\Lm\pi R}.
\ee

This result is independent of the level number $p$.
However, it is not a general feature of our SUSY breaking 
mechanism.
It depends on the choice of the interaction $W_{\rm int}$.
If we choose $W_{\rm int}=-h\Phi\Phi_{\rm m}^{2}$ as an 
example, 
we will obtain a different result that 
$\dms_{p}$ becomes larger as $p$ increases, 
just like the result in Ref.\cite{MSSS}.

\renewcommand{\dy}{\!\!{\rm d}y}
\section{Soft SUSY breaking terms in 3D effective theory} 
\label{3D-eff}
In this section, 
we discuss how various soft SUSY breaking terms 
in the three-dimensional effective theory 
are induced in our framework. 

Firstly, we discuss a multi-linear scalar coupling, a 
generalization of 
the so-called A-term.
Such a ``generalized A-term'' is generated from the following 
superpotential 
term in the bulk theory
\begin{eqnarray}
\Sat &\!\!\!=&\!\!\! \int \dX{4} 
\left.
\frac{\cF\left(\Phi(X,\tht)/\Mf\right)}{\Mf^{\Nm-3}} 
\Phi_{i_1}(X,\tht) \cdots \Phi_{i_\Nm}(X,\tht)
\right|_{\tht^2} 
+ {\rm h.c.} \label{general-Wint} \\
&\!\!\!\supset&\!\!\! \int \dX{4} 
\frac{\cF'(\Acl(y)/\Mf)\Fcl(y)}{\Mf^{\Nm-2}} 
A_{i_1}(X) \cdots A_{i_\Nm}(X) + {\rm h.c.}, \\ 
&\!\!\!\supset&\!\!\! 2\left\{\int \dy 
\frac{\cF'(\Acl(y)/\Mf)\Fcl(y)}{2^{\Nm/2}
\Mf^{\Nm-2}} \bmR{i_1}{0}(y) \cdots \bmR{i_\Nm}{0}(y) \right\}
\int \dx{3} \amR{i_1}{0}(x) \cdots \amR{i_\Nm}{0}(x), 
\nonumber\\
&\!\!\!&\!\!\! \label{directA}
\end{eqnarray}
where $\Mf$ is the fundamental mass scale of the 
four-dimensional bulk theory, 
$\cF(\phi)$ is a dimensionless holomorphic function of 
$\phi$, 
and $\Phi_i$ $(i=1,\cdots,N_{\rm m})$ are chiral matter 
superfields,
\be
 \Phi_i=A_i+\sqrt{2}\tht\Psi_i+\tht^2 F_i.
\ee
The equation of motion for $F_{\rm cl}$ is given by 
\be
 \Fcl(y)\equiv -\left.\frac{\del W^\ast}{\del 
A^\ast}\right|_{A=\Acl(y)}. 
\ee
Note that the the superpotential term Eq.(\ref{general-Wint}) 
is a generalization of Eq.(\ref{Wint}).
In Eq.(\ref{directA}), we used the following Kaluza-Klein (KK) 
mode expansions,
\begin{eqnarray}
A_i(X) &\!\!=&\!\! \frac{1}{\sqrt{2}} 
(A_{{\rm R}i}(X) +i A_{{\rm I}i}(X)), \nonumber\\
A_{{\rm R}i}(X) &\!\! = &\!\! \sum_p \bmR{i}{p}(y)\amR{i}{p}(x), 
\qquad 
A_{{\rm I}i}(X) 
=
 \sum_p \bmI{i}{p}(y)\amI{i}{p}(x). 
\label{Ai_expansion}
\end{eqnarray}

When the number $\Nm$ of the matter fields is three, the 
$y$-integral 
in Eq.(\ref{directA}) becomes an A-parameter 
in three-dimensional effective theory. 
When $\Nm=2$, $\Sat$ in Eq.(\ref{directA}) becomes a 
so-called 
B-term and 
also includes the following Yukawa interactions 
\bea
\label{WeylA}
 \Sat &\!\!\!\supset&\!\!\! -\int\dX{4} 
\cF'\left(\frac{\Acl(y)}{\Mf}\right)
 \Psi(X)\left(A_i(X)\Psi_j(X)+\Psi_i(X)A_j(X)\right)+{\rm h.c.} 
\\
 &\!\!\!\supset&\!\!\! 
 \int\dx{3}\left\{ \gefb{ij}\psng(x)\amR{i}{0}(x)\ps{1}{j,0}(x) 
 +\gefb{ji}\psng(x)\amR{j}{0}(x)\ps{1}{i,0}(x)\right\}, 
\nonumber\\
 &\!\!\!&\!\!\!\label{overA}
\eea
where the effective coupling constant $\gefb{ij}$ is defined by 
\be
 \gefb{ij}=
 -\frac{1}{\sqrt{2}}\int\dy\cF'\left(\frac{\Acl(y)}{\Mf}\right)
 \fng(y)\bmR{i}{0}(y)\f{1}{j,0}(y),  \label{gefb}
\ee
and the Weyl fermion $\Psi_i(X)$ 
is rewritten by Majorana fermions $\Psi^{(1,2)}_i(X)$ and 
mode-expanded just like $\Psi(X)$ in Eqs.(\ref{fluc_fields}) 
and 
(\ref{eq:fermion_mode_decomp})
\bea
 \Psi_i(X)&\!\!=&\!\!\frac{1}{\sqrt{2}}(\Psi^{(1)}_i(X)+i\Psi^{(2)}_i
(X)), 
 \nonumber\\
 \Psi^{(1)}_i(X)&\!\!=&\!\!\sum_p\f{1}{i,p}(y)\ps{1}{i,p}(x), 
 \quad 
 \Psi^{(2)}_i(X)
 =
 \sum_p\f{2}{i,p}(y)\ps{2}{i,p}(x).
\eea

We now turn to the squared scalar masses.
They are generated from the following K\"{a}hler potential 
term 
\bea
 \Ssc &\!\!\!=&\!\!\! \int \dX{4} 
 \left.\cG\left(\frac{\Phi(X,\tht)}{\Mf},\frac{\bar{\Phi}(X,\bar{
\tht})}
 {\Mf}\right) 
 \bar{\Phi}_i(X,\bar{\tht}) 
 \Phi_j(X,\tht)\right|_{\theta^2\bar{\theta}^2}, \label{S_scalar} 
\\
 &\!\!\!\supset&\!\!\! \left\{\int\dy \frac{\cG_{\phi\bar{\phi}}
(\Acl(y)/\Mf)\Fcl^2(y)}
 {2\Mf^2} 
 \bmR{i}{0}(y)\bmR{j}{0}(y)\right\} \int\dx{3} \amR{i}{0}(x)\amR{j}
{0}(x), 
 \label{universal_scalar_mass}
\end{eqnarray}
where $\cG(\phi,\bar{\phi})$ is a real function and 
$\cG_{\phi\bar{\phi}}(\Acl/\Mf)\equiv(\del_{\phi} 
\del_{\bar{\phi}}\cG)
(\Acl/\Mf,\Acl/\Mf)$. 
We used the mode expansion Eq.(\ref{Ai_expansion}) 
and the fact that $\Fcl(y)$ is real.

$\Ssc$ also involves the following interactions 
\bea
 \Ssc &\!\!\!\supset&\!\!\! -\int \dX{4} 
 \frac{\cG_{\phi\bar{\phi}}(\Acl(y)/\Mf)\Fcl(y)}{\Mf^2} 
 \Psi(X)\left(A^*_i(X) \Psi_j(X)
 +\Psi_i(X)A^*_j(X)\right) +{\rm h.c.}, \nonumber\\
 &\!\!\!&\!\!\! \label{wscalar} \\
 &\!\!\!\supset&\!\!\! \int\dx{3} 
\left\{\gefs{ij}\psng(x)\amR{i}{0}(x)\ps{1}{j,0}(x)  
 +\gefs{ji} \psng(x)\amR{j}{0}(x)\ps{1}{i,0}(x)\right\}, 
\nonumber\\
 &\!\!\!&\!\!\!\label{oversclr}
\eea
where the effective coupling constant $\gefs{ij}$ is defined by 
\be
 \gefs{ij}=
 -\frac{1}{\sqrt{2}\Mf^2}\int\dy\cG_{\phi\bar{\phi}}
 \left(\frac{\Acl(y)}{\Mf}\right)
 \Fcl(y)\fng(y)\bmR{i}{0}(y)\f{1}{j,0}(y). \label{gefs}
\ee

It should be noted that the squared scalar mass terms and 
the so-called B-term are indistinguishable in three 
dimensions, 
because fields in three dimensions are real. 
We emphasize that the low-energy theorem (see 
Appendix~\ref{LET-mixing})
\be
 \gef{ij}=\frac{\dms_{ij}}{\sqrt{2}f}
\ee
relates the mass-splitting 
$\dms_{ij}$ 
and the Yukawa coupling constant $\gef{ij}$, which in general 
receive 
contributions from various terms like Eq.(\ref{directA}) 
($\Nm=2$) 
and Eq.(\ref{universal_scalar_mass}) for $\dms_{ij}$, 
and Eq.(\ref{overA}) and Eq.(\ref{oversclr}) for $\gef{ij}$, 
respectively.

Finally, we consider the gauge supermultiplets.
The gaugino mass has a contribution from the following 
non-minimal gauge kinetic term in the bulk theory. 
\begin{eqnarray}
\label{gaugino}
\Sgg &\!\!=&\!\! \int \dX{4} 
\left.\cH\left(\frac{\Phi(X,\tht)}{\Mf}\right) 
W^{\alpha}(X,\tht)W_{\alpha}(X,\tht) \right|_{\tht^2} + {\rm 
h.c.}, \\
\label{gaugino1}
&\!\!\supset&\!\! -\int \dX{4} \frac{\cH'(\Acl(y)/\Mf)F_{{\rm 
cl}}(y)}{\Mf} 
(\lambda^2(X) + \bar{\lambda}^2(X)),
\end{eqnarray}
Where $\cH(\phi)$ is a holomorphic function of $\phi$, and 
$W_{\alpha}$ is a field strength superfield and can be written 
by 
component fields as
\be
 W_{\alpha}=-i\lm_{\alpha}+\left\{\delta_{\alpha}^{\ \beta}D
 -\frac{i}{2}\left(\sigma^{\mu}\bar{\sigma}^{\nu}\right)_{\alpha}
^{\ \beta}
 V_{\mu\nu}\right\}\tht_{\beta}
 +\tht^2\sigma_{\alpha\dot{\alpha}}^{\mu}\del_{\mu}\bar{\lm}^{
\dot{\alpha}},
\ee
in the Wess-Zumino gauge.
The spinor $\lm$ is a gaugino field and $V_{\mu\nu}$ is a 
field strength of 
the gauge field, and $D$ is an auxiliary field.

\section{Soft SUSY breaking terms in 4D effective theory} 
\label{4D-eff}
In this section, we discuss the soft SUSY breaking terms 
in four-dimensional effective theory reduced from 
the five-dimensional ${\cal N}=1$ theory. 
We will use the superfield formalism proposed in 
Ref.\cite{AGW} that 
keeps only the four-dimensional ${\cal N}=1$ supersymmetry 
manifest. 
The four-dimensional SUSY that we keep manifest is the one 
preserved by 
our wall in the limit of $R\to\infty$, and we call it 
$Q^{(1)}$-SUSY.
We do not specify a mechanism to form our wall 
and the other wall.
We assume the existence of a pair of chiral supermultiplets 
$\Phi=A+\sqrt{2}\tht\Psi+\tht^2 F$ and 
$\Phi^c=A^c+\sqrt{2}\tht\Psi^c+\tht^2 F^c$, 
forming a hypermultiplet of the four-dimensional 
${\cal N}=2$ supersymmetry. 
Their F-components have non-trivial classical values 
$\Fcl(y)$ and 
$\Fcl^c(y)$. 
In the following, the background field configuration $\Acl(y)$, 
$\Acl^{c}(y)$, and $\Fcl(y)$, $\Fcl^{c}(y)$ 
are assumed to be real for simplicity.
In this section, 
$X$ and $x$ represent five- and four-dimensional 
coordinates respectively, 
and $y$ denotes the coordinate of the extra dimension.

The relevant term to generate the generalized A-term is 
\begin{eqnarray}
\label{4dAterm}
\Sat &\!\!=&\!\! \int \dX{5} 
\left.\frac{\cF(\Phi(X,\tht)/\Mf^{3/2},\Phi^c(X,\tht)/\Mf^{3/
2})}
{\Mf^{(3\Nm-8)/2}} 
\Phi_1(X,\tht) \cdots \Phi_{\Nm}(X,\tht)\right|_{\tht^2} 
+ {\rm h.c.} \\
&\!\!\supset&\!\! \int \dX{5} 
\frac{\del\cF(y)\Fcl(y)+\del^c\cF(y)\Fcl^c(y)}{\Mf^{(3\Nm-5)
/2}}
A_1(X) \cdots A_{\Nm}(X) + {\rm h.c.}, \\
&\!\!\supset&\!\! \left\{\int\dy 
\frac{\del\cF(y)\Fcl(y)+\del^c\cF(y)\Fcl^c(y)}{\Mf^{(3\Nm-5)
/2}}
b_{1,0}(y) \cdots b_{\Nm,0}(y)\right\} 
\int \dx{4} a_{1,0}(x) \cdots a_{\Nm,0}(x) 
+ {\rm h.c.}, \nonumber\\
&\!\!\!&\!\!\! \label{direct4dA}
\end{eqnarray}
where $\Mf$ is the fundamental mass scale of the 
five-dimensional bulk theory. 
Note that the superfields $\Phi$, $\Phi^c$ and $\Phi_i$ in 
five dimensions 
have mass-dimension 3/2.
$\cF(\phi,\phi^c)$ is a holomorphic function of $\phi$ and 
$\phi^c$, 
and
\be
 \del\cF(y)\equiv(\del_{\phi}\cF)\left(\frac{\Acl(y)}{\Mf^{3/2}}
,
 \frac{\Acl^c(y)}{\Mf^{3/2}}\right), \;\;\;
 \del^c\cF(y)\equiv(\del_{\phi^c}\cF)\left(\frac{\Acl(y)}{\Mf^{
3/2}},
 \frac{\Acl^c(y)}{\Mf^{3/2}}\right).
\ee
In Eq.(\ref{direct4dA}), we used the following mode expansion, 
\be
 A_i(X)=\sum_p b_{i,p}(y)a_{i,p}(x). 
\ee
%
The $y$-integral in Eq.~(\ref{direct4dA}) is 
a generalized A-parameter in four-dimensional effective 
theory. 
{}For example, the usual A- and B-parameters have 
contributions 
from $N=3$ and $N=2$ respectively 
\begin{eqnarray}
A_{ijk} &\!\!=&\!\! \int\dy 
\frac{\del\cF(y)\Fcl(y)+\del^c\cF(y)\Fcl^c(y)}
{\Mf^2}
b_{i,0}(y) b_{j,0}(y) b_{k,0}(y), \\
-B_{ij} \mu &\!\!=&\!\! \int\dy 
\frac{\del\cF(y)\Fcl(y)+\del^c\cF(y)\Fcl^c(y)}
{\sqrt{\Mf}} 
 b_{i,0}(y) b_{j,0}(y),
 \label{eq:Bterm}
\end{eqnarray}
where $\mu$ is the so-called $\mu$-parameter.

When $\Nm=2$, $\Sat$ in Eq.(\ref{4dAterm}) also includes 
the following 
Yukawa interaction 
\bea
 \label{4dA}
 \Sat &\!\!\supset&\!\! -\int\dX{5} 
 \frac{\del\cF(y)\Psi(X)+\del^c\cF(y)\Psi^c(X)}{\sqrt{\Mf}}
 \left(A_{i}(X)\Psi_{j}(X)+\Psi_{i}(X)A_{j}(X)\right) + {\rm h.c.} \\
 &\!\!\supset&\!\! \int\dx{4} 
\left\{\gefb{ij}\psng(x)a_{i,0}(x)\psi_{j,0}(x)
 +\gefb{ji}\psng(x)a_{j,0}(x)\psi_{i,0}(x) \right\}+{\rm h.c.},
 \nonumber\\
 &\!\!\!&\!\!\! \label{over4dA}
\end{eqnarray}
\be
 \gefb{ij}=-\int\dy 
 \frac{\del\cF(y)\fng(y)+\del^c\cF(y)\fng^c(y)}{\sqrt{\Mf}}
 b_{i,0}(y)f_{j,0}(y).
\ee
Here we used the mode expansion of $\Psi(X)$, $\Psi^c(X)$ 
and $\Psi_i(X)$, 
\be
 \left(\begin{array}{c} \Psi(X) \\ \Psi^c(X) \end{array}\right)
 =\sum_p \left(\begin{array}{c}f_{p}(y) \\ 
f_{p}^c(y)\end{array}\right)
 \psi_{p}(x),
\ee
\be
 \Psi_i(X)=\sum_p f_{i,p}(y)\psi_{i,p}(x). 
\ee
In general, the NG fermion $\psng(x)$ is contained in both 
$\Psi(X)$ and 
$\Psi^c(X)$ with mode functions $\fng(y)$ and $\fng^c(y)$, 
respectively. 
By definition, $\fng(y)$ and $\fng^c(y)$ have their support 
mainly 
on the other wall.

Next, we discuss the squared scalar masses. 
The squared scalar masses get contributions from the term 
with a real function $\cG(\phi,\phi^c,\bar{\phi},\bar{\phi}^c)$, 
\begin{eqnarray}
\label{4dscalar}
\Ssc &\!\!=&\!\! \int \dX{5} 
\cG\left(\frac{\Phi(X,\tht)}{\Mf^{3/2}},
\frac{\Phi^c(X,\tht)}{\Mf^{3/2}},
\frac{\bar{\Phi}(X,\bar{\tht})}{\Mf^{3/2}},
\frac{\bar{\Phi}^c(X,\bar{\tht})}{\Mf^{3/2}}\right)
\bar{\Phi}_i(X,\bar{\tht})\Phi_j(X,\tht)|_{\theta^2\bar{\theta}^2}
, \\
&\!\!\supset&\!\! \int \dX{5} \frac{\tilde{\cG}(y)}{M^3_*} 
A_i^*(X) A_j(X), \\ 
\label{soverlap}
&\!\!\supset&\!\! \left\{\int\dy \frac{\tilde{\cG}(y)}{M^3_*} 
b_{i,0}^*(y)b_{j,0}(y) \right\}
\int \dx{4} a_{i,0}^{*}(x)a_{j,0}(x),
\end{eqnarray}
where functions $\tilde{\cG}(y)$, $\cG_{\phi\bar{\phi}}(y), 
\cG_{\phi\bar{\phi}^c}(y), \cdots$ are defined by 
\be
 \tilde{\cG}(y)\equiv \cG_{\phi\bar{\phi}}(y)\Fcl^2(y)
 +\cG_{\phi\bar{\phi}^c}(y)\Fcl(y)\Fcl^c(y)
 +\cG_{\phi^c\bar{\phi}}(y)\Fcl^c(y)\Fcl(y)
 +\cG_{\phi^c\bar{\phi}^c}(y)\left(\Fcl^c(y)\right)^2, 
\ee
\be
 \cG_{\phi\bar{\phi}}(y)\equiv(\del_{\phi}\del_{\bar{\phi}}\cG)
 \left(\frac{\Acl(y)}{\Mf^{3/2}},\frac{\Acl^c(y)}{\Mf^{3/2}},
 \frac{\Acl^*(y)}{\Mf^{3/2}},\frac{\Acl^{c*}(y)}{\Mf^{3/2}}\right)
, 
 \quad \cdots. 
\ee

%

The following Yukawa interactions are also contained in 
$\Ssc$ in Eq.(\ref{4dscalar}), 
%
\bea
 \Ssc &\!\!\supset&\!\! -\int\dX{5} 
 \frac{\cG_{\phi\bar{\phi}}(y)\Fcl(y)+\cG_{\phi\bar{\phi}^c}(y)
 \Fcl^c(y)}
 {\Mf^3} 
 \Psi(x)\left( A^*_i(X) \Psi_j(X)+\Psi_i(X)A^*_j(X)\right) + {\rm 
h.c.}, 
 \nonumber\\
 &\!\!\!&\!\!\! \label{4dscalarlet} \\
 &\!\!\supset&\!\!\int\dx{4}\left\{ \gefs{ij}\psng(x)a_{i,0}^*(x)
\psi_{j,0}(x)
 +\gefs{ji}\psng(x)a_{j,0}^*(x)\psi_{i,0}(x)\right\} + {\rm h.c.},
 \nonumber\\
 &\!\!\!&\!\!\! \label{overscalar}
\eea
where the effective Yukawa coupling $\gefs{ij}$ is defined by 
\be
 \gefs{ij}\equiv 
-\frac{1}{\Mf^3}\int\dy\left(\cG_{\phi\bar{\phi}}(y)\Fcl(y)
 +\cG_{\phi\bar{\phi}^c}(y)\Fcl^c(y)\right)\fng(y)b_{i,0}^*(y)f_{j,
0}(y).
 \label{eq:yukawa_coupling}
\ee
Just like the three-dimensional case, the low-energy 
theorem 
\be
 \gef{ij}=-\frac{\dms_{ij}}{f} \label{gef-GTR4d}
\ee
is valid in four dimensions (See Eq.(\ref{GTR_4Dchiral}) 
in Appendix~\ref{4D_LET_chiral}.), 
where $f$ is the order parameter of the SUSY breaking. 
Both the mass-splittings $\Delta m_{ij}^2$ and  the effective 
couplings 
$g_{{\rm eff} ij}$ are the sum of contributions from various 
terms. 
However, the squared mass terms and the B-term are 
distinguished by chirality of scalar fields in four dimensions, 
unlike the three-dimensional case. 
Therefore the low-energy theorem should be valid separately 
for the squared 
mass terms and the B-term relating to the effective 
couplings of the 
corresponding chirality. 

{}Finally, we consider the gaugino mass.
Note that the gauge supermultiplet in five-dimensional ${\cal 
N}=1$ theory 
contains two gauginos in a four-dimensional ${\cal N}=1$ 
sense. 
However, since we are interested only in a four-dimensional 
${\cal N}=1$ SUSY, 
$Q^{(1)}$-SUSY, we will consider only $\lm_0(x)$, which is 
a $Q^{(1)}$-superpartner of the gauge field $v_{\mu\nu,0}(x)$, 
as the gaugino. 
The gaugino mass has a contribution from the term
with a holomorphic function $\cH(\phi,\phi^c)$ of $\phi$ and 
$\phi^c$
\begin{equation}
\label{5dgaugino}
 \Sgg = \int\dX{5} \left.
 \cH\left(\frac{\Phi(X,\tht)}{\Mf^{3/2}},
 \frac{\Phi^c(X,\tht)}{\Mf^{3/2}}
 \right)
 W^\alpha(X,\tht) W_\alpha(X,\tht)\right|_{\theta^2}.
\end{equation}
Performing the mode expansion of the gauge supermultiplet,
\bea
 V_{\mu\nu}(X)&\!\!=&\!\!\sum_{p}b_{v,p}(y)v_{\mu\nu,p}(x), 
 \quad 
 \lm(X)
 =
 \sum_p f_{\lm,p}(y)\lm_{p}(x),
\eea 
we can see $\Sgg$ contains the following term 
\begin{equation}
 \Sgg \supset 
 \left\{\int\dy \frac{\del\cH(y)\Fcl(y)+\del^c\cH(y)\Fcl^c(y)}
{\Mf^{3/2}} 
 \left(f_{\lm,0}(y)\right)^2\right\} 
 \int\dx{4} \left(\lm_0(x)\right)^2, \label{gaugino_mass1}
\end{equation}
where 
\be
 \del\cH(y)\equiv 
(\del_{\phi}\cH)\left(\frac{\Acl(y)}{\Mf^{3/2}},
 \frac{\Acl^c(y)}{\Mf^{3/2}}\right), \;\;\;
 \del^c\cH(y)\equiv 
(\del_{\phi^c}\cH)\left(\frac{\Acl(y)}{\Mf^{3/2}},
 \frac{\Acl^c(y)}{\Mf^{3/2}}\right).
\ee
Eq.(\ref{gaugino_mass1}) contributes to the mass of the 
gaugino $\lm_0(x)$.
In order to obtain the gaugino mass-eigenvalue itself, we have 
to take account 
of the derivative term in the extra dimension $y$, and define 
a differential 
operator ${\cal O}_\lm$ like the left-hand side of 
Eq.(\ref{fermion_mode_eq}). 
However, it is very difficult to find eigenvalues of ${\cal 
O}_\lm$ generally. 
Therefore the single-wall approximation explained 
in section~\ref{single-wall_ap} 
is quite a powerful method to estimate $m_\lm$, 
thanks to the low-energy theorem. 

The term (\ref{5dgaugino}) also includes the following 
interaction
\bea
 \label{4dlet}
 \Sgg &\!\!\supset&\!\! \int\dX{5} \frac{1}{\sqrt{2}\Mf^{3/2}} 
 \lambda(X)\sigma^\mu \bar{\sigma}^\nu 
 \left\{\del\cH(y)\Psi(X)+\del^c\cH(y)\Psi^c(X)\right\}
 V_{\mu\nu}(X) 
 + {\rm h.c.} \\
 &\!\!\supset&\!\! \hef{}\int \dx{4} \lm_{0}(x) \sigma^\mu 
\bar{\sigma}^\nu 
 \psng(x) v_{\mu\nu,0}(x) + {\rm h.c.}, \label{goverlap}
\eea
where the effective coupling constant $\hef{}$ is defined by 
\be
 \hef{}=\int \dy \frac{1}{\sqrt{2} \Mf^{3/2}} 
 f_{\lm,0}(y)\left(\del\cH(y)\fng(y)+\del^c\cH(y)\fng^c(y)\right) 
 b_{v,0}(y).
 \label{hef-gauge}
\ee 
This effective coupling constant is related to the 
mass-splitting of 
the gauge supermultiplet, which equals the gaugino mass 
$m_\lm$, 
and the order parameter of the SUSY breaking $f$ by the 
low-energy theorem 
\be
 \hef{}=\frac{m_{\lm}}{\sqrt{2}f}. \label{4d-GTR}
\ee
This theorem is derived in Appendix~\ref{4D_LET_gauge}. 

Using Eq.(\ref{4d-GTR}), we can estimate the gaugino mass 
$m_\lm$ 
by calculating the effective coupling constant $\hef{}$ 
in Eq.(\ref{hef-gauge}).
For example, if the gauge supermultiplet lives in the bulk, 
zero-mode wave functions $f_{\lm,0}(y)$ and $b_{v,0}(y)$ 
become constant: 
$1/\sqrt{2\pi R}$ in the single-wall approximation.
Thus the gaugino mass is estimated as
\be
\label{4dgaugino2}
 m_\lm=\frac{1}{4\sqrt{2}\pi \Mf^{3/2}R}
 \int\dy\left(\del\cH(y)\fng(y)+\del^c\cH(y)\fng^c(y)\right).
\ee

\section{Phenomenological implications} 
\label{phenomeno-imp}

\begin{figure}[t]
\leavevmode
\epsfysize=6cm
\centerline{\epsfbox{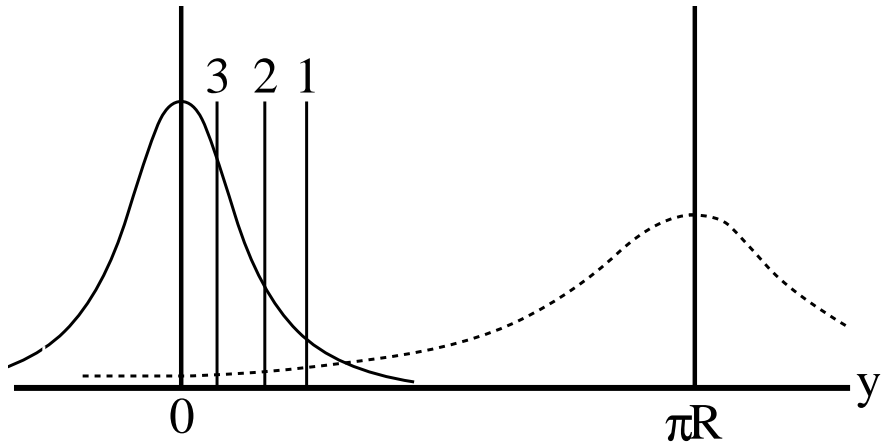}}
\caption{Schematic picture of the location of the matter 
fields. 
1, 2 and 3 represents the location of the first, second and 
third generation of the matter fields. 
The solid line denotes the Higgs wave function and 
the dotted line denotes the wave function of NG fermion. 
}
\label{location}
\end{figure}

Here the qualitative phenomenological features 
in our framework will briefly be discussed. 
It is well known that 
information of fermion masses and mixings can be translated 
into 
the locations of the wave functions for matter fields 
in extra dimensions \cite{DvaliShifman,KT,AS,MS,GP}. 
Yukawa coupling in five dimensions is written as 
\begin{eqnarray}
\label{yukawa}
S_{\rm Yukawa}=\int \dX{5} \left( 
\frac{y^u_{ij}}{\sqrt{M}} Q_i(X,\tht) U^c_j(X,\tht) H_2(X,\tht) + 
\frac{y^d_{ij}}{\sqrt{M}} Q_i(X,\tht) D^c_j(X,\tht) H_1(X,\tht) 
\right. 
\nonumber \\
\left. \left.
+ \frac{y^l_{ij}}{\sqrt{M}} L_i(X,\tht) E^c_j(X,\tht) H_1(X,\tht) 
\right)\right|_{\theta^2} 
+ {\rm h.c.}
\end{eqnarray}
where $y^u_{ij}, y^d_{ij}$ and $y^l_{ij}$ are dimensionless 
Yukawa coupling 
constants for up-type quark, down-type quark and 
charged lepton sector of order unity, respectively. 
The fundamental mass scale of the five-dimensional bulk 
theory is denoted 
by $M$. 
Notice that additional contributions to 
Yukawa coupling Eq.(\ref{yukawa}) 
come from terms like Eq.(\ref{4dAterm}). 
If we consider $M$ as the gravitational scale $M_*$, 
these contributions are subleading compared to 
Eq.(\ref{yukawa}). 
On the other hand, 
if $M$ happens to be the scale of the wall, such as 
$\Lambda$, 
these contributions will be comparable to Eq.(\ref{yukawa}). 
Here we simply write down Eq.(\ref{yukawa}), since an 
analysis of fermion 
masses and mixings is not the main point of this paper. 
Performing the mode expansion for each matter 
supermultiplet, 
we obtain, for example, up-type Yukawa coupling from 
Eq.~(\ref{yukawa}),
\begin{equation}
\label{YUKAWA}
S_{\rm Yukawa}\supset \left\{\int\dy \frac{y^u_{ij}}{\sqrt{M}} 
f_{Q_i,0}(y) f_{U^c_j,0}(y) b_{H_2,0}(y)\right\} 
\int\dx{4} q_{i,0}(x) u^c_{j,0}(x) h_{2,0}(x), 
\end{equation}
where $q_{i,0}(x)$ and $u^c_{j,0}(x)$ are massless fields 
of fermionic components of 
$Q_i(X,\tht)$ and $U^c_j(X,\tht)$, and $h_{2,0}(x)$ is a  
massless field 
of a bosonic component of $H_2(X,\tht)$, respectively. 
$f_{Q_i,0}(y)$, $f_{U^c_j,0}(y)$ and $b_{H_2,0}(y)$ are 
corresponding 
mode functions.
The effective Yukawa coupling in four dimensions is 
the $y$-integral part of Eq.(\ref{YUKAWA}). 
Fermion masses and mixings are determined 
by the overlap integral between Higgs and matter fields. 
For example, the hierarchy of Yukawa coupling is generated 
by shifting the locations of the wave functions 
slightly generation by generation \cite{KT}. 
These shifts are easily achieved by introducing 
five-dimensional mass terms 
in a generation-dependent way. 
The fermion masses exhibit a hierarchy $m_1<m_2<m_3$, 
where $m_{1,2,3}$ denote masses of the first, second and 
third generation of matter fermions respectively. 
Therefore we can naively expect 
that the locations of the wave functions of matter fields 
$y_i (i=1,2,3)$ become $y_1>y_2>y_3(>0)$ 
if the Higgs is localized\footnote{
Of course, we can also take $y_1<y_2<y_3(<0)$ to realize the 
fermion mass 
hierarchy, but these two cases are equivalent 
since the extra dimension is compactified.
} around $y=0$ as shown in Fig.~\ref{location}. 

%
%

In the two-wall background configuration, SUSY is broken 
and fermion and 
sfermion masses split. 
Even though it is difficult to solve mass-eigenvalues directly, 
we can calculate the mass-splitting in each supermultiplet 
 thanks to the low-energy theorem Eqs.(\ref{gef-GTR4d}) and 
 (\ref{4d-GTR}). 
The overlap integral in Eq.(\ref{overscalar}) among the chiral 
supermultiplets localized on our wall and the NG fermion 
localized 
on the other wall determines the mass-splitting and hence 
sfermion masses. 
Thus the mass-eigenvalue of the sfermion becomes larger as 
the location 
is closer to the other wall.

Before estimating the sfermion mass spectrum, 
we comment on various scales in our theory. 
There are four typical scales in our theory: 
the five-dimensional Planck scale $M_*$, 
the compactification scale $(2 \pi R )^{-1}$, 
the inverse width of the wall $\Lambda$ and 
the inverse width $a$ of zero-mode wave functions. 
In order for our setup to make sense, we had better keep 
the following relation among these scales 
\begin{equation}
\label{scale}
M_* > a > \Lambda > ( 2 \pi R )^{-1} > 5000~{\rm TeV}. 
\label{eq:scales_inequality}
\end{equation}
The inequality $a>\Lm$ comes from the requirement that our 
wall must have 
enough width to trap matter modes.
The last constraint is required to suppress flavor changing 
neutral 
currents mediated by Kaluza-Klein gauge bosons \cite{DPQ}. 
If we consider the flat background metric, $M_*$ and $R$ are 
related by 
the relation $M_{\rm pl}^2=(2\pi R)M_*^3$ where $M_{\rm pl}$ 
is 
the four-dimensional Planck scale.
Thus the above constraint gives the lower bound for $M_*$, 
that is 
\be
 M_*=\left(\frac{M_{\rm pl}^2}{2\pi R}\right)^{1/3} > 
 \left(M_{\rm pl}^2\times 5000\mbox{TeV}\right)^{1/3}
 \simeq 8 \times 10^{14}\mbox{GeV}. 
\ee

Now, 
we would like to make a rough estimation of 
the gravity at the tree level 
by applying the results in section 4 and considering the scale 
$M$ 
as the five-dimensional Planck scale $M_*$. 
Let us start with the sfermion masses. 
We recall that 
the interaction Eq.(\ref{4dscalar}) gives Yukawa coupling 
Eq.(\ref{eq:yukawa_coupling}) 
\be
g_{{\rm eff}ij} = -\frac{1}{M_*^3} 
\int\dy F_{{\rm cl}}(y) f_{{\rm NG}}(y) b_{i,0}^*(y) f_{j,0}(y) 
\quad (i,j = 1,2,3), 
\ee
where we assumed $\cG=\bar{\Phi}\Phi/M_*^3$ 
for simplicity. 
On the other hand, 
the low-energy theorem for the chiral supermultiplet 
(\ref{GTR_4Dchiral}) is 
\be
g_{{\rm eff}} = - \frac{\Delta m^2}{f}.
\ee
Assuming the fermion masses are small, we find that the 
sfermion masses are 
given by 
\bea
(\tilde{m}^2)_{ij} &\!\!=&\!\! \frac{f}{M_*^3} 
\int\dy \Fcl(y) f_{{\rm NG}}(y) b_{i,0}^*(y) f_{j,0}(y) 
\label{sfermion}
\eea
The classical configuration $\Acl(y)$ is 
approximately linear 
in $y$ in the vicinity of the wall, and constant away from the 
wall. 
Correspondingly 
we can approximate $F_{{\rm cl}}(y)$ by a Gaussian function 
and the wave function of NG fermion by an exponential 
function, 
if we consider a large distance between two walls. 
We also adopt the Gaussian approximation for the zero-mode 
wave 
functions of the matter fields 
\be
b_{i,0}(y) \simeq f_{i,0}(y) \simeq N_a \exp[-a^2(y-y_i)^2], 
\ee
where $y_i$ is a location of the matter field 
and $a$ represents a typical inverse width and 
$N_a$ is a normalization constant of 
the zero-mode wave function for matter fields. 
Thus we obtain sfermion masses 
\bea
(\tilde{m}^2)_{ij} &\!\!\simeq&\!\! N_a^2 \frac{f}{M_*^3} 
\int\dy~(\Lambda^{5/2}e^{-\Lambda^2 y^2})~(\sqrt{\Lambda}
e^{-\Lambda(\pi R-y)})~e^{-a^2(y-y_i)^2}~e^{-a^2(y-y_j)^2} \\
%
\label{sfermionmass}
 &\!\!\simeq &\!\! \frac{f}{\sqrt{2}} 
\left( \frac{\Lambda}{M_*} \right)^3 
\frac{{\rm Erf}~[\sqrt{2} \pi Ra]}
{{\rm Erf}~[\pi Ra]} e^{-\Lambda \pi R}
{\rm exp}\left[ -\frac{a^2}{2}(y_i-y_j)^2 \right], \label{til-m_ij}
\eea
where 
the error function Erf [$x$] is defined as 
${\rm Erf}~[x] \equiv \frac{2}{\sqrt{\pi}} \int^x_0 dy e^{-y^2}$ 
and 
the normalization constants 
$N_a^2 = \frac{a}{\sqrt{\pi}
{\rm Erf}~[\pi Ra]}$ 
are substituted. 
The approximation $2 \pi R \gg y_i$ is used in the second line. 
One can see that the sfermion mass matrix is determined 
by only the relative difference of the coordinates 
where the matter fields are localized. 
The dependence of the distance between the location of the matter and 
the other wall is subleading.
Using the typical example in Ref.\cite{KT} which well 
reproduces 
the fermion mass hierarchy and their mixings 
$y_1 \sim 3.05 M_*^{-1}, 
y_2 \sim 2.29 M_*^{-1}, 
y_3 \sim 0.36 M_*^{-1}$,
and diagonalizing the sfermion mass matrix, 
we obtain the following results. 
If we consider the case $a \sim M_*$, 
the overlap between the wave functions of the different 
generations 
is small because the width of the wave function is small. 
Hence the hierarchy of the sfermion masses is at most one order of 
magnitude. 
On the other hand, if we consider the case $a \sim 0.1M_*$, 
the overlap between the wave functions of the different 
generations 
is larger, and 
all the matrix elements of sfermion mass squared matrix are 
nearly equal. 
In this case, 
the rank of the sfermion mass matrix is reduced, 
then 
the sfermion mass becomes ${\cal O}(10 {\rm TeV}), {\cal 
O}(1 {\rm TeV})$ 
and ${\cal O}(100 {\rm GeV})$. 
Although this result looks like the decoupling solution 
\cite{decoupling} for FCNC problem, it has a 
 mixing among the generations too large to be a viable 
solution 
 for the FCNC problem. 
Since this result is an artifact of our rough approximation, 
we expect that a more realistic sfermion masses can be 
obtained, 
if we take account of flexibility of the model, such as the 
location and 
shape of the wave functions. 

We now turn to the case of gaugino. 
Let us first consider the case that the gauge supermultiplet 
lives in the 
bulk. 
Eqs.(\ref{eq:yukawa_coupling}) and (\ref{hef-gauge}) show that 
the overlap integral for the chiral supermultiplet receives 
an exponential suppression but that for the gauge 
supermultiplet does not. 
The gaugino tends to be heavier than the sfermions in this 
case. 
There are three ways to avoid this situation. 
One of them is to tune the numerical coefficient of the term 
Eq.(\ref{5dgaugino}) to be small. 
The second way is to localize 
the gauge supermultiplet on our wall. 
The third way is to assume that the function $\del\cH$ and 
$\del^c\cH$ 
in Eq.(\ref{4dgaugino2}) have profiles which are localized on 
our wall. 
Then, even if the gauge supermultiplet lives in the bulk, 
the gaugino mass is suppressed 
because of the suppression of the overlap with the NG 
fermion localized 
on the other wall. 

Next we consider the case that the gauge supermultiplet is 
localized 
on the wall. 
We also assume that 
 the wave function of the zero mode of gauge supermultiplet 
is 
 Gaussian 
\be
f_{\lambda_1,0}(y) = b_{v,0}(y) \sim {\rm exp} (-a^2 y^2). 
\ee
Since Eq.(\ref{hef-gauge}) gives  the gaugino mass 
through the low-energy theorem Eq.(\ref{4d-GTR}), 
we find by taking the limit $\pi R \gg \Lambda/(4a^2)$ 
\bea
m_{\lambda} &\!\!=&\!\!  \frac{f}{M_*^{3/2}} 
\int\dy f_{\lambda_1,0}(y) f_{{\rm NG}}(y)b_{v,0}(y), \\
&\!\!\simeq&\!\! f \left( \frac{\Lambda}{2M_*^3} 
\right)^{1/2}
{\rm exp} \left( -\Lambda \pi R + \frac{\Lambda^2}{8a^2} 
\right) 
\frac{{\rm Erf}~[\sqrt{2} \pi Ra]}
{{\rm Erf}~[\pi Ra]}, \label{m_lmd}
\eea
where we assumed that the gauge kinetic function is 
${\cal H} = \Phi/M^{3/2}$. 
Requiring $|m_{\lambda}| \sim {\cal O}(100$GeV), 
$\tilde{m}^2 \sim {\cal O}({\rm TeV}^2)$ and 
${\rm exp}[-\frac{a^2}{2}(y_i-y_j)^2] \simeq {\cal O}(0.1) \sim 
{\cal O}(1)$, 
we obtain 
\be
\Lambda \sim 10^{1.6 \sim 2}\left({M_* \over {\rm 
GeV}}\right)^{3/5} 
{\rm GeV}. 
\ee
Taking Eq.(\ref{scale}) into account, we obtain the bounds for 
$M_*$ and $\Lambda$ 
\bea
\label{bound1}
8 \times 10^{14}~{\rm GeV} < M_* < 3 \times 10^{16}~{\rm 
GeV}, \\
\label{bound2}
9 \times 10^{10}~{\rm GeV} < \Lambda < 3 \times 
10^{11}~{\rm GeV}. 
\eea
SUSY breaking scale can be obtained from the gaugino mass 
as 
\be
\sqrt{f} \sim 2 \times 10^{11}~{\rm GeV},
\ee
where we have used $\Lambda \sim 10^{11}~{\rm GeV}$ and 
$M_* \sim 10^{16}~{\rm GeV}$. 
SUSY breaking scale is comparable to 
that of the gravity mediation $\sqrt{f} \sim 
10^{10\sim11}$~GeV.

{}Finally, some comments are in order. 
The above Eqs.(\ref{til-m_ij}) and (\ref{m_lmd}) include only effects of 
light modes at tree level of gravitational interaction. 
We would like to compare these gravity mediated contributions 
with those induced by coexisting walls ($M=\Lm$). 
The bilinear term of the five-dimensional gravitino has a coefficient 
of order $c_g \Fcl(y)/M_*^{3/2}$ in the case of the gravity mediation, 
and of order $c_w \Fcl(y)/\Lm^{3/2}$ in the case of coexisting walls, 
where $c_g$ and $c_w$ are numerical constants.
As long as we have no information about the fundamental 
theory, 
we cannot calculate these constants $c_g, c_w$ in the effective theory. 
Taking the ratio of these contributions, we obtain 
\be
\frac{\mbox{non-gravity}}{{\rm gravity}} \sim 
\frac{c_w}{c_g}\left( \frac{M_*}{\Lambda} \right)^{3/2} 
\sim \frac{c_w}{c_g}\left( \frac{10^{16}}{10^{11}} \right)^{3/2} 
\sim \frac{c_w}{c_g}\cdot 10^{7.5}. 
\ee
If $c_w/c_g > 10^{-7}$, the gravity mediated contribution is smaller 
than the non-gravity mediated contribution. 
If $c_w/c_g < 10^{-8}$, the gravity mediated contribution is larger 
than the non-gravity mediated contribution. 

The second comment is on the proton stability in our 
framework. 
In the ``fat brane" approach, 
it is well known that the operators which are relevant to 
the proton decay are exponentially suppressed 
by separating the quark wave functions from the lepton wave 
functions 
\cite{AS}. 
This mechanism also works in our model. 
Noticing that the fifth dimension is compactified on a circle, 
it is sufficient for the wave functions of the quark and the 
lepton to be 
localized on the opposite side with respect to 
the plane $y=0$ where the Higgs field is localized.
This relative location is required to reproduce the quark and 
lepton masses. 
Let us suppose that the distance 
between the locations of the quark and the lepton is $r$. 
Then, the dimension five operators are suppressed by 
$e^{-(ar)^2}/M_* = \frac{1}{M_P}\frac{M_P}{M_*}e^{-(ar)^2}$, 
where $M_P$ is the Planck scale in four dimensions. 
To keep the proton stable enough as required by 
experiments, 
$\frac{M_P}{M_*}e^{-(ar)^2} \sim 10^{-7}$ is needed. 
This constraint is indeed satisfied 
if we take $M_* \sim 10^{16}$ GeV 
and $ar \sim {\cal O}(\mbox{5 - 6})$, and is consistent with 
Eq.(\ref{eq:scales_inequality}). 
Thus, the proton decay process is easily suppressed in our 
framework.

\section{Discussion} \label{discuss}

In this paper, we proposed a simple SUSY breaking 
mechanism 
in the brane world scenario.
The essence of our mechanism is just the coexistence of 
two different kinds of 
BPS domain walls at finite distance. 
Our mechanism needs no messenger fields nor complicated 
SUSY breaking sector 
on any of the walls.
The low-energy theorem provides a powerful method to 
estimate 
the boson-fermion mass-splitting.
Namely, the mass-splitting can be estimated by calculating an 
overlap integral 
of the mode functions for matter fields and the NG fermion.
Matter fields are localized on our wall by definition. 
On the other hand, since the supersymmetry approximately 
preserved 
on our wall is broken due to the existence of the other wall, 
the corresponding NG fermion is localized on the other wall.
Thus the mass-splitting induced in the effective theory is 
exponentially 
suppressed compared to the fundamental scale $\Lm$.
This is the generic feature of our mechanism.

Now let us discuss several further issues.

As mentioned below Eq.(\ref{GTR-gef}), 
the order parameter of the SUSY breaking $f$ 
is equal to the square root of the energy density of the wall 
$\sqrt{V_0}$. 
From the three-dimensional point of view, the fundamental 
theory 
is an ${\cal N}=2$ SUSY theory with $Q^{(1)}$- and 
$Q^{(2)}$-SUSYs.
In general when a BPS domain wall exist, a half of the bulk 
SUSY, for example, 
$Q^{(2)}$-SUSY, is broken.
In such a case, an order parameter of the SUSY breaking 
$f_2$ is equal to 
the square root of the energy density of the domain wall 
$\sqrt{V_0}$. 
However, if there is another BPS domain wall that breaks the 
other 
half of the bulk SUSY, $Q^{(1)}$-SUSY, 
there is another order parameter of the SUSY breaking $f_1$ 
and its square 
is expected to be equal to the energy density of the 
additional wall.
In the model discussed in section~\ref{SUSY-br-coexist}, 
these two order parameters $f_1$ and $f_2$ are equal to each 
other. 
This is because the two walls are symmetric in this model.
However in the case when our wall and the other wall are not 
symmetric, 
two order parameters $f_1$ and $f_2$ can have different 
values.
In Appendix~\ref{V0-f1f2}, we discuss the possibility of such 
an asymmetric 
wall-configuration and the relation among $f_1$, $f_2$ and 
$V_0$ and central 
charge of the SUSY algebra.

If we try to construct a realistic model in our SUSY breaking 
mechanism, 
a fundamental bulk theory, which has a five-dimensional 
${\cal N}=1$ SUSY, 
must have BPS domain walls.
Since such a higher dimensional SUSY restricts 
the form of the superpotential severely, it is not easy to 
construct a BPS 
domain wall configuration.
However, a  BPS domain wall has been constructed in a 
four-dimensional 
${\cal N}=2$ SUSY non-linear sigma model\cite{townsend}. 
Since the nonlinear sigma model can be obtained from the 
 ${\cal N}=1$ five-dimensional theory, this BPS domain wall 
can be regarded 
as a BPS domain wall that we desire. 
It is more difficult to obtain non-BPS configuration of two 
walls. 

Our mechanism can be extended to higher dimensional cases 
straightforwardly. 
In such cases, our four-dimensional world is on various kinds 
of 
topological defects, 
such as vortices or intersections of domain walls in six 
dimensions, 
monopoles in seven dimensions, etc.
Many higher dimensional theories have BPS configurations of 
these defects.
Thus all we need for our mechanism is a stable non-BPS 
configuration 
corresponding to the coexistence of two or more BPS 
topological defects 
that preserve different parts of the bulk SUSY. 
We can always use the low-energy theorem like 
Eqs.(\ref{gef-GTR4d}) 
and (\ref{4d-GTR}) irrespective of the dimension of the bulk 
theory,
in order to estimate the mass-splittings between bosons and 
fermions.

As a future work, we will investigate our SUSY breaking 
mechanism 
in the non-trivial metric like the Randall-Sundrum background 
\cite{RS}. 
To achieve this goal, we need to overcome the technical 
complexity 
of dealing with the five-dimensional supergravity. 
Besides, when we introduce the gravity, the size of the fifth 
dimension 
$2\pi R$ becomes a dynamical variable. 
In the model discussed in section~\ref{SUSY-br-coexist}, for 
example, 
the force between our wall and the other wall is repulsive. 
Thus the two-wall configuration Eq.(\ref{A_classical}) 
becomes unstable 
($2\pi R$ goes to infinity) when the gravity is considered. 
So we must implement an extra mechanism to stabilize the 
two-wall 
configuration not only topologically but also under the gravity.


\renewcommand{\thesubsection}{Acknowledgments}
\subsection{}

One of the authors (N.S.) is indebted to useful discussions 
with 
Kiwoon Choi, Takeo Inami, Ken-ichi Izawa, Martin Schmaltz, Tsutomu 
Yanagida, and 
Masahiro Yamaguchi. 
One of the authors (N.M.) thanks to a discussion with Hitoshi 
Murayama. 
This work is supported in part by Grant-in-Aid for Scientific 
Research from the Ministry of Education, Culture, Sports, 
Science and 
Technology,Japan, priority area(\#707) ``Supersymmetry and 
unified theory
of elementary particles" and No.13640269. 
N.M.,Y.S.~and R.S.~are supported 
by the Japan Society for the Promotion of Science for Young 
Scientists 
(No.08557, No.10113 and No.6665).

\renewcommand{\thesubsection}{\thesection.\arabic{subsection}}

\appendix

\section{Low-energy theorem in three dimensions} 
\label{LET-mixing}
In this appendix, we will review 
the low-energy theorem for the SUSY breaking briefly, and 
apply it to our mechanism.

\subsection{SUSY Goldberger-Treiman relation} 
\label{LET-in-3D}
In general, when the supersymmetry is spontaneously broken, 
a massless fermion called the Nambu-Goldstone (NG) 
fermion $\psng(x)$ appears 
in the theory.
It shows up in the supercurrent $J^{m}_{\alpha}(x)$ as 
follows\cite{clark} 
\be
 J^{m}_{\alpha}=\sqrt{2}if\left(\gm{m}\psng\right)_{\alpha}
 +J^{m}_{\phi,\alpha}+\cdots, \label{super_cc_3D}
\ee
where $f$ is the order parameter of the SUSY breaking and 
the abbreviation denotes higher order terms for $\psng(x)$. 
$J^{m}_{\phi,\alpha}(x)$ is the supercurrent for matter fields 
$\phi=(a,\psi)$ where $a(x)$ and $\psi(x)$ are a real scalar 
and 
a Majorana spinor fields respectively, 
\be
 J^{m}_{\phi,\alpha}=\left(\gm{n}\gm{m}\psi\right)_{\alpha}\del_
{n}a+\cdots.
\ee

In the low-energy effective Lagrangian, there is a Yukawa 
coupling as follows.
\be
 \cL_{\rm Yukawa}=\gef{}a\psi\psng.
\ee
Here the effective coupling constant $\gef{}$ is related to 
the mass-splitting between the boson and the fermion 
$\dms\equiv m_{a}^{2}-m_{\psi}^{2}$ and 
the order parameter $f$ by\cite{clark}
\be
 \gef{}=\frac{\dms}{\sqrt{2}f}. \label{GTR-hef}
\ee

This is the supersymmetric analog of the Goldberger-Treiman 
relation. 


\subsection{Superpartners and mass-eigenstates} 
\label{sp-me}
When SUSY is broken, a superpartner of a fermionic 
mass-eigenstate is not 
always a mass-eigenstate.
In such a case, we should extend the formula 
Eq.(\ref{GTR-hef}) to 
more generic form.

Let us denote fermionic mass-eigenstates as 
$\psi_{1}, \psi_{2}, \cdots, \psi_{N}$, and their bosonic 
superpartners as 
$a_{1}, a_{2}, \cdots, a_{N}$.
The bosonic mass-eigenstates $\amt{1}, \amt{2}, \cdots, 
\amt{N}$ are related 
to $a_{1}, a_{2}, \cdots, a_{N}$ by 
\be
 \left(\begin{array}{c} a_{1} \\ a_{2} \\ \vdots \\ a_{N} 
\end{array}\right)
 =V\left(\begin{array}{c} \amt{1} \\ \amt{2} \\ \vdots \\ 
\amt{N} \end{array}
 \right), \label{V-definition}
\ee
where $V$ is an $N\times N$ unitary mixing matrix.

In this case, Eq.(\ref{GTR-hef}) is generalized to 
\be
 \gef{i,j}=\frac{1}{\sqrt{2}f}\left(\dMs\right)_{j,i}, 
\label{GTR-hef_ij}
\ee
where $\gef{i,j}$ are Yukawa coupling constants: 
\be
 \cL_{\rm Yukawa}=\sum_{i,j}\gef{i,j}a_{i}\psi_{j}\psng, 
\ee
and $\dMs$ is an $N\times N$ matrix defined by 
\be
 \dMs\equiv V\left(\begin{array}{ccc}m_{a_{1}}^{2} & & \\ & 
\ddots & \\ 
 & & m_{a_{N}}^{2} \end{array}\right)
 -\left(\begin{array}{ccc}m_{\psi_{1}}^{2} & & \\ & \ddots & \\
 & & m_{\psi_{N}}^{2} \end{array}\right)V.
\ee

\subsection{Application to our model} \label{application}
To apply the above low-energy theorem to our mechanism of 
the SUSY breaking, 
we should interpret the four-(five-)dimensional bulk theory 
as a three-(four-)dimensional theory involving infinite 
Kaluza-Klein modes. 
To illustrate this, let us discuss the low-energy theorem by 
using 
the model Eq.(\ref{Logn}) in the four-dimensional bulk as an 
example.

\subsubsection{Three-dimensional super-transformation}
The superpartner of $\ps{1}{p}(x)$ for $Q^{(1)}$-SUSY can be 
read off from the four-dimensional super-transformation, 
\be
 \delta_{\xi}A(X)=\sqrt{2}\xi\Psi(X),
\ee
where $\xi$ is a Weyl spinor which parametrizes the 
super-transformation.
By expanding the four-dimensional fields $A$ and $\Psi$ to 
infinite 
Kaluza-Klein modes like Eqs.(\ref{fluc_fields}), 
(\ref{eq:boson_mode_decomp}), 
(\ref{eq:fermion_mode_decomp}), multiplying $\f{1}{p}(y)$ and 
integrating in terms of $y$, we can obtain the 
three-dimensional 
super-transformation.
\be
 \delta_{\zeta}\sum_{q}\left\{\left(\int\dy\f{1}{p}(y)\bR{q}(y) 
 \right)
 \ar{q}(x)\right\}=\zeta\ps{1}{p}(x),
\ee
where $\zeta$ denotes the parameter of 
$Q^{(1)}$-transformation, which is 
a three-dimensional Majorana spinor.

Thus the superpartner of $\ps{1}{p}(x)$ for $Q^{(1)}$-SUSY, 
$a^{(1)}_{p}(x)$, is a linear combination of infinite 
mass-eigenmodes.
\be
 a^{(1)}_{p}(x)=\sum_{q}\left(\int\dy\f{1}{p}(y)\bR{q}(y)\right)\ar{
q}(x).
 \label{a1p_decomp}
\ee
This is because $Q^{(1)}$-SUSY is broken by the background 
$\Acl(y)$. 
When the distance between the walls is infinite, 
$Q^{(1)}$-SUSY is recovered 
and $a^{(1)}_p(x)$ becomes a mass-eigenmode.
In this case, $Q^{(2)}$-SUSY is also recovered and 
$a^{(2)}_p(x)$, which is 
a superpartner of the mass-eigenmode $\ps{2}{p}(x)$, 
becomes a mass-eigenmode. 
Since the fields $a^{(1)}_p(x)$ and $a^{(2)}_p(x)$ are 
degenerate, 
they maximally mix when the wall distance is finite. 
For example, 
\be
 \left(\begin{array}{c} \ar{0} \\ \ar{1} \end{array}\right)
 \simeq\frac{1}{\sqrt{2}}\left(\begin{array}{cc} 1 & -1 \\ 1 & 1 
\end{array}
 \right)\left(\begin{array}{c} a^{(1)}_0 \\ a^{(2)}_0 
\end{array}\right), 
\ee
that is, 
\be
 a^{(1)}_{0}\simeq\frac{1}{\sqrt{2}}\left(\ar{0}+\ar{1}\right).
 \label{a0-ap}
\ee
This can be directly obtained from Eq.(\ref{a1p_decomp}) by 
setting $p=0$.

Strictly speaking, $a^{(1)}_{0}(x)$ has slight but non-zero 
components 
of heavier fields $\ar{p}(x)$~($p\geq 2$). 
However these components become negligibly small as $p$ 
increases.
Thus by introducing a cutoff $N$ for the Kaluza-Klein level 
and setting it large enough, 
we can apply the formula Eq.(\ref{GTR-hef_ij}) to our case. 
The mixing matrix $V$ in Eq.(\ref{V-definition}) can be read 
off 
from Eq.(\ref{a1p_decomp}) as follows. 
\be
 V_{p,q}=\int\dy\f{1}{p}(y)\bR{q}(y).
\ee

\subsubsection{Derivation of the formula Eq.(\ref{GTR-gef})}
Here we will derive the formula Eq.(\ref{GTR-gef}), as an 
example. 
Since the effective coupling constant $\gef{}$ in 
Eq.(\ref{effthry}) is 
$\gef{1,0}$ in the notation here, 
it is related to the element $(\dMs)_{0,1}$ 
according to Eq.(\ref{GTR-hef_ij})  
\be
 \left(\dMs\right)_{0,1}=V_{0,1}\mR{1},
\quad 
 V_{0,1}=\int\dy\f{1}{0}(y)\bR{1}(y)=k\frac{C_{0}}{\CR{1}},
\ee
where normalization factors $C_{0}$ and $\CR{1}$ are defined 
by 
Eq.(\ref{bosonR_mode_fnc}) and Eq.(\ref{fermion_mode_fnc}), 
and 
\bea
 C_{0}&\!\!=&\!\!\left(\int\dy\left\{\dn\left(\frac{\Lm 
y}{k},k\right)
 +k\cn\left(\frac{\Lm y}{k},k\right)\right\}^{2}\right)^{-1/2}
 \nonumber\\
 &\!\!=&\!\!\left(\int\dy\left\{(1+k^{2})-2k^{2}\sn^{2}
 \left(\frac{\Lm y}{k},k\right)
 \right\}\right)^{-1/2}=\left(V_{0}\frac{g^{2}k^{2}}{\Lm^{4}}\right
)^{-1/2}
 =\frac{\Lm^{2}}{fgk}.
\eea
Here we used Eq.(\ref{eq:vacuum_energy}) and the relation 
$V_{0}=f^{2}$.
Then we find  the low-energy theorem Eq.(\ref{GTR-hef_ij}) 
using Eq.(\ref{geff}) 
\be
 \frac{1}{\sqrt{2}f}\left(\dMs\right)_{0,1}=\frac{V_{0,1}\mR{1}^{2}
}{\sqrt{2}f}
 =\frac{1}{\sqrt{2}f}\cdot k\frac{C_{0}}{\CR{1}}\cdot
 \frac{1-k^{2}}{k^{2}}\Lm^{2}
 =\frac{g}{\sqrt{2}}\frac{C_{0}^{2}}{\CR{1}}(1-k^{2})=\gef{}.
\ee
When the distance between the walls is large, 
$V_{0,1}\simeq 1/\sqrt{2}$ and we obtain Eq.(\ref{GTR-gef}). 

In the above calculation, we assumed that the normalization 
factors 
$C_0$, $\CR{0}$ and $\CR{1}$ are all positive.
In fact, we can calculate the boson-fermion mass-splittings 
including their sign, 
irrespective of the sign conventions of these normalization 
factors.
Next, we will show this fact.

\subsubsection{Unambiguity of the sign of the mass-splitting}
Firstly, we should note that the sign of the normalization 
factor of 
the NG fermion $C_{0}$ is determined by the convention of 
the sign of 
the order parameter $f$.

The supercurrent in Eq.(\ref{super_cc_3D}) can be obtained 
from 
that of the bulk theory, 
\be
 J^{\mu}_{\alpha}=\sqrt{2}(\sigma^{\nu}\bar{\sigma}^{\mu}\Psi)
_{\alpha}
 \del_{\nu}A^{\ast}-i\sqrt{2}(\sigma^{\mu}\bar{\Psi})_{\alpha}
 \frac{\del W^{\ast}}{\del A^{\ast}}. \label{super_cc_4D}
\ee

We define the three-dimensional currents 
$J^{(1)m}_{\alpha}(x)$ 
and $J^{(2)m}_{\alpha}(x)$ as follows.
\be
 \int\dy J^{m}_{\alpha}(X)=\frac{1}{\sqrt{2}}\left(
 J^{(1)m}_{\alpha}(x)+iJ^{(2)m}_{\alpha}(x)\right), 
 \label{3D-supercurrents}
\ee
where $J^{(1)m}_{\alpha}(x)$ and $J^{(2)m}_{\alpha}(x)$ are 
three-dimensional 
Majorana currents.

By substituting the mode expansion of fields: 
\bea
 A(x,y)&\!\!=&\!\!\Acl(y)+\frac{1}{\sqrt{2}}\left\{
 \sum_{p}\bR{p}(y)\ar{p}(x)+i\sum_{p}\bI{p}(y)\ai{p}(x)\right\},
 \nonumber\\
 \Psi(x,y)&\!\!=&\!\!\frac{1}{\sqrt{2}}\left\{
 \sum_{p}\f{1}{p}(y)\ps{1}{p}(x)+i\sum_{p}\f{2}{p}(y)\ps{2}{p}(x)
 \right\},
\eea
into $J^{(1)m}_{\alpha}(x)$, we can obtain the 
three-dimensional supercurrent 
for $Q^{(1)}$-SUSY 
\bea
 J^{(1)m}(x)&\!\!=&\!\!\sqrt{2}i\left\{\int\dy\f{2}{0}(y)\left(
 \del_{y}\Acl(y)-\frac{\Lm^{2}}{g}\cos\left(\frac{g}{\Lm}\Acl(y)
\right)\right)
 \right\}\gm{m}\ps{2}{0}(x) \nonumber\\
 &&\!\!+\sum_{p,q}V_{p,q}\gm{n}\gm{m}\ps{1}{p}(x)\del_{n}\ar{q}(
x)
 +\cdots.
\eea

Comparing this to Eq.(\ref{super_cc_3D}), we can see that the 
order parameter 
of the SUSY breaking $f$ is expressed by 
\be
 f=\int\dy\f{2}{0}(y)\left(\del_{y}\Acl(y)-\frac{\Lm^{2}}{g}
 \cos\left(
 \frac{g}{\Lm}\Acl(y)\right)\right)=\frac{\Lm^{2}}{gkC_{0}}.
\ee
Thus if we take a convention of $f>0$, 
the normalization factor $C_{0}$ is set to be positive.

Noticing that $(\dMs)_{p,q}=V_{p,q}(\mR{q}^{2}-m_{p}^{2})$, 
we obtain the following formula from Eq.(\ref{GTR-hef_ij})
\bea
 \mR{q}^{2}-m_{p}^{2}&\!\!=&\!\!\sqrt{2}f\frac{\gef{q,p}}{V_{p,q}} 
\nonumber\\
 &\!\!=&\!\!\frac{\sqrt{2}\Lm^{2}}{gkC_{0}}
 \frac{\int\dy\bR{q}(y)\f{1}{p}(y)\f{2}{0}(y)}
 {\int\dy\f{1}{p}(y)\bR{q}(y)}. \label{sign_unambiguity_formula}
\eea

Therefore we can calculate the mass-splitting 
$\mR{q}^{2}-m_{p}^{2}$ 
including its sign, irrespective of the sign conventions 
of the normalization factors.

\subsubsection{Estimation in the single-wall approximation} 
\label{E-SWA}
Finally, we comment on the estimation of the mass-splitting 
in the single-wall approximation (SWA).
When we estimate the boson-fermion mass-splitting in SWA, 
we often approximate the bosonic mode function 
by that of its fermionic superpartner 
in the calculation of the overlap integral.
This means that we estimate the following effective coupling 
as $\gef{ij}$ 
in Eq.(\ref{GTR-hef_ij}).
\be
 \cL_{\rm Yukawa}=\gef{p}^{\rm 
(SWA)}a^{(1)}_p\ps{1}{p}\ps{2}{0}+\cdots.
\ee

As mentioned above, the superpartner $a^{(1)}_p(x)$ of the 
fermionic mass 
eigenmode 
$\ps{1}{p}(x)$ is a linear combination of mainly two bosonic 
mass-eigenmodes 
\be
 a^{(1)}_p(x)\simeq\frac{1}{\sqrt{2}}(\ar{2p}(x)+\ar{2p+1}(x)).
\ee
Thus corresponding mode function $b^{(1)}_p(y)$ is 
\be
 b^{(1)}_p(y)\simeq\frac{1}{\sqrt{2}}(\bR{2p}(y)+\bR{2p+1}(y)).
\ee

Then by using $b^{(1)}_p(y)$, which is well-approximated by 
$\f{1}{p}(y)$, 
as a bosonic mode function, the formula 
Eq.(\ref{sign_unambiguity_formula}) 
becomes 
\bea
 \sqrt{2}f\gef{p}^{\rm (SWA)}&\!\!=&\!\!
 \sqrt{2}f\int\dy b^{(1)}_p(y)\f{1}{p}(y)\f{2}{0}(y) \nonumber\\
 &\!\!\simeq &\!\! f\int\dy (\bR{2p}(y)+\bR{2p+1}(y))\f{1}{p}(y)
\f{2}{0}(y)
 \nonumber\\
 &\!\!\simeq &\!\! f\frac{1}
{\sqrt{2}}\left(\frac{\gef{2p,p}}{V_{2p,p}}
 +\frac{\gef{2p+1,p}}{V_{2p+1,p}}\right) \nonumber\\
 &\!\!=&\!\!\frac{1}{2}\left\{(\mR{2p}^2-m_p^2)+(\mR{2p+1}^2-
m_p^2)\right\} 
 \nonumber\\
 &\!\!=&\!\! \frac{\mR{2p}^2+\mR{2p+1}^2}{2}-m_p^2,
\eea
where we used the fact that $V_{2p,p}\simeq V_{2p+1,p}\simeq 
1/\sqrt{2}$, 
and the coupling constant $\gef{p}^{\rm (SWA)}$ is defined by 
\be
 \gef{p}^{\rm (SWA)}\equiv \int\dy 
b^{(1)}_p(y)\f{1}{p}(y)\f{2}{0}(y)
 \simeq \int\dy \left(\f{1}{p}(y)\right)^2\f{2}{0}(y).
\ee

Therefore what we can estimate in the single-wall 
approximation is 
the difference between a fermionic mass and an average of 
squared masses 
of its bosonic superpartners.

\section{Low-energy theorem in four dimensions} 
\label{LETin4D}
In this appendix, we derive the low-energy theorem for chiral 
and 
gauge supermultiplets in four dimensions.
We will follow the procedure in Ref.\cite{clark}.

\subsection{Low-energy theorem for Chiral supermultiplets} 
\label{4D_LET_chiral}
Let us denote one-particle state of a scalar boson with the 
mass $m_{\rm B}$ 
and the momentum $\pB$ as $|\pB\rangle$, and that of a 
spin 1/2 fermion 
with the mass $m_{\rm F}$ and the momentum $\pF$ as 
$|\pF\rangle$, 
which form a chiral supermultiplet.
We perform the Lorentz decomposition of a matrix element 
for the supercurrent $J^{\mu}_{\alpha}(x)$ between these 
states.
\bea
 \langle \pB |J^{\mu}_{\alpha}(0)|\pF\rangle &\!\!=&\!\! 
 [A_1(q^2)q^{\mu}+A_2(q^2)k^{\mu}+
 A_3(q^2)\sigma^{\mu}\bar{\sigma}^{\nu}q_{\nu}]_{\alpha}^{\ \ 
\beta}
 \chF{\beta}(\pF) \nonumber\\
 &\!\!\!&\!\!+[A_4(q^2)\sigma^{\mu}+A_5(q^2)q^{\mu}\sigma^{
\nu}q_{\nu}
 +A_6(q^2)k^{\mu}\sigma^{\nu}q_{\nu}]_{\alpha\dot{\beta}}
 \bchF{\dot{\beta}}(\pF), 
 \label{chiral_decomp}
\eea
where $q^{\mu}\equiv \pB^{\mu}-\pF^{\mu}$ and 
$k^{\mu}\equiv \pB^{\mu}+\pF^{\mu}$. 
The spinors $\chF{}(\pF)$ and $\bchF{}(\pF)$ obey the 
following equations 
\bea
 \sigma\cdot \pF\bchF{}(\pF)&\!\!=&\!\! m_{\rm 
F}\chF{}(\pF), \quad
 \bar{\sigma}\cdot \pF\chF{}(\pF)
 =
  m_{\rm F}\bchF{}(\pF).
\eea

Conservation of the supercurrent leads to a relation among 
the form factors as
\be
 q^2\left[A_1(q^2)-A_3(q^2)\right]=\dms A_2(q^2), 
\label{A1-A3}
\ee
where $\dms\equiv m_{\rm B}^2-m_{\rm F}^2$ is a 
mass-splitting 
between the boson and the fermion.

To discuss S-matrix elements, we define an NG fermion 
source 
$j^{\rm NG}_{\alpha}(x)$ 
by using the NG fermion field $\psng(x)$ as 
\be
 j^{\rm 
NG}_{\alpha}(x)=-i\sigma^{\mu}_{\alpha\dot{\alpha}}\del_{\mu}
 \bar{\psi}_{\rm NG}^{\dot{\alpha}}(x).
\ee
Its matrix element between the boson and the fermion states 
is decomposed as 
\be
 \langle \pB|j^{\rm NG}_{\alpha}(0)|\pF\rangle=
 B_1(q^2)\chF{\alpha}(\pF)+B_2(q^2)q\cdot\sigma_{\alpha\dot
{\alpha}}
 \bchF{\dot{\alpha}}(\pF), \label{jNG_decomp}
\ee
and thus
\be
 \langle \pB|\bar{\psi}_{\rm NG}^{\dot{\alpha}}(0)|\pF\rangle=
 -\frac{B_1(q^2)}{q^2}q\cdot\bar{\sigma}^{\dot{\alpha}\alpha}
 \chF{\alpha}(\pF)
 +B_2(q^2)\bchF{\dot{\alpha}}(\pF). \label{NGbar_decomp}
\ee

Since the combination 
$J^{\mu}_{\alpha}-\sqrt{2}if\sigma^{\mu}_{\alpha\dot{\alpha}}
\bar{\psi}_{\rm NG}^{\dot{\alpha}}$ 
has vanishing matrix element between the vacuum and the 
single NG fermion 
state, all the form factors of 
$\langle 
\pB|J^{\mu}_{\alpha}-\sqrt{2}if\sigma^{\mu}_{\alpha\dot{\alpha
}}
\bar{\psi}_{\rm NG}^{\dot{\alpha}} |\pF\rangle$ 
are regular as $q^2\to 0$. 
Then comparing Eqs.(\ref{chiral_decomp}) and 
(\ref{NGbar_decomp}), 
we can see that the form factor $A_3(q^2)$ is singular at 
$q^2=0$ 
unless $B_1(0)$ is zero. 
\be
 \lim_{q^2\to 0}q^2 A_3(q^2)=-\sqrt{2}ifB_1(0).
\ee
Substituting it into Eq.(\ref{A1-A3}) with the limit $q^2\to 0$, 
we obtain 
\be
 \sqrt{2}ifB_1(0)=\dms A_2(0). \label{B1-A2}
\ee

To relate the form factor $B_1(0)$ to an effective 
 coupling constant of the NG fermion with 
 the boson and the fermion forming a chiral supermultiplet, 
we evaluate a transition amplitude between the in-state 
$|q;\pF\rangle_{\rm in}$ 
and the out-state $|\pB\rangle_{\rm out}$.
This S-matrix element can be expressed by using an 
effective 
interaction Lagrangian 
$\cL_{\rm int}$ as
\bea
 \mbox{}_{\rm out}\langle \pB|q;\pF\rangle_{\rm in}&\!\!=&\!\!
 \mbox{}_{\rm I}\langle\pB |e^{i\int\dx{4}\cL_{\rm 
int}(x)}|\pF\rangle_{\rm I}
 \nonumber\\
 &\!\!\simeq&\!\! i(2\pi)^4\delta^4(\pB-\pF-q)\:
 \mbox{}_{\rm I}\langle \pB|\cL_{\rm int}(0)|q;\pF\rangle_{\rm 
I}, 
 \label{B-NG_F}
\eea
where $|\pB\rangle_{\rm I}$ and $|q;\pF\rangle_{\rm I}$ 
denote 
states in the interaction picture.

On the other hand, using the LSZ reduction formula, it can 
also be written as 
\bea
 \mbox{}_{\rm in}\langle\pB |q;\pF\rangle_{\rm out}&\!\!=&\!\!
 -i(2\pi)^4\delta^4(\pB-\pF-q)\chng(q)q_{\mu}\sigma^{\mu}
 \mbox{}_{\rm I}\langle\pB |\bpsng(0)|\pF\rangle_{\rm I} 
\nonumber\\
 &\!\!\!&\!\!-i(2\pi)^4\delta^4(\pB-\pF-q)\bchng(q)q_{\mu}
 \bar{\sigma}^{\mu}
 \mbox{}_{\rm I}\langle\pB |\psng(0)|\pF\rangle_{\rm I},
\eea
where $\chng(q)$ and $\bchng(q)$ are the NG fermion 
spinors. 
Since we do not need to distinguish the interaction picture 
and the 
Heisenberg picture for one-particle states, we drop the 
subscript I 
for one-particle states in the following. 
We obtain a relation between matrix elements of the 
interaction Lagrangian 
and the NG fermion field 
\be
 \langle\pB |\cL_{\rm int}(0)|q;\pF\rangle_{\rm I}=
 -\chng(q)q_{\mu}\sigma^{\mu}
 \langle\pB |\bpsng(0)|\pF\rangle
 -\bchng(q)q_{\mu}\bar{\sigma}^{\mu}
 \langle\pB |\psng(0)|\pF\rangle. \label{ME-of-Lint}
\ee

At soft NG fermion limit $q^{\mu}\to 0$, the S-matrix 
element Eq.(\ref{B-NG_F}) 
should be expressible by the following nonderivative 
interaction terms 
in the effective Lagrangian \cite{lee-wu}
\be
 \cL_{\rm int}=\gef{}a^*\psi\psng+{\rm h.c.}+\cdots,
\ee
where $a$ is a complex scalar field and $\psi$ is a 
two-component Weyl spinor  
field, which create or annihilate the states $|\pB\rangle$ and 
$|\pF\rangle$ respectively. 
So its matrix element is written as 
\be
 \langle \pB|\cL_{\rm int}(0)|q;\pF\rangle_{\rm I}=
 \gef{}\chng(q)\chF{}(\pF)+\gef{}\bchng(q)\bchF{}(\pF).
 \label{ME-of-Lint_gef}
\ee

Substituting Eq.(\ref{NGbar_decomp}) into 
Eq.(\ref{ME-of-Lint}), and 
comparing it with Eq.(\ref{ME-of-Lint_gef}) gives a relation 
between $B_1(0)$ and $\gef{}$ 
\be
 B_1(0)=-\gef{}.
\ee
Thus Eq.(\ref{B1-A2}) becomes 
 the supersymmetric analog of the Goldberger-Treiman 
relation 
\be
 \sqrt{2}\gef{}f=i(m_{\rm B}^2-m_{\rm F}^2)A_2(0). \label{gf-A2}
\ee

Noting that the supercurrent takes the form 
\be
 J^{\mu}_{\alpha}=\sqrt{2}if\sigma^{\mu}_{\alpha\dot{\alpha}}
 \bar{\psi}_{\rm NG}^{\dot{\alpha}}
 +\sqrt{2}\left(\sigma^{\nu}\bar{\sigma}^{\mu}\psi\right)_{\alpha}
 \del_{\nu}a^*+\cdots, 
\ee
and substituting it into the left-hand-side of 
Eq.(\ref{chiral_decomp}) 
with the limit $q^2\to 0$, 
we can determine the value of the form factor $A_2(0)$ as 
\be
 A_2(0)=\sqrt{2}i.
\ee
Thus we obtain 
 the low-energy theorem for the chiral supermultiplets from 
Eq.(\ref{gf-A2})
\be
 \gef{}=-\frac{m_{\rm B}^2-m_{\rm F}^2}{f}. 
\label{GTR_4Dchiral}
\ee


\subsection{Low-energy theorem for Gauge supermultiplets}
\label{4D_LET_gauge}
Next we derive the low-energy theorem for gauge 
supermultiplets.
As the case of chiral supermultiplets, 
we consider the Lorentz decomposition of the matrix element 
for the supercurrent $J^{\mu}_{\alpha}(x)$ between 
one-particle state of 
the gauge boson $|\pB\rangle$ with the mass $m_{\rm B}$ 
and the momentum 
$\pB$, 
and that of the gaugino $|\pF\rangle$ with the mass $m_{\rm 
F}$ and the 
momentum $\pF$
\begin{equation}
\begin{array}{l}
\langle\pB |J^{\mu}_\alpha(0)| \pF \rangle \vspace{1mm}
\\
 \ \ \ = \ \epsilon^{\ast}_\nu(\pB) \bigl[
   A_1(q^2) q^\nu q^\mu 
 + A_2(q^2) q^\nu k^\mu 
 + A_3(q^2) q^\nu \sigma^\mu \bar{\sigma}^\rho q_\rho 
 + A_4(q^2) \eta^{\mu\nu}
 + A_5(q^2)\sigma^\nu \bar{\sigma}^\mu
 \bigr]_\alpha^{\ \ \beta} \chF{\beta}(\pF) \vspace{1mm}\\
 \ \ \ \ + \ \epsilon^\ast_\nu (\pB) \bigl[
   A_6(q^2)q^\nu \sigma^\mu
 + A_7(q^2)q^\nu q^\mu \sigma^\rho q_\rho
 + A_8(q^2)q^\nu k^\mu \sigma^\rho q_\rho
 + A_9(q^2)\eta^{\mu \nu}\sigma^\rho q_\rho  \vspace{1mm} 
\\
 \hspace{5cm}
+ A_{10}(q^2)q^\mu \sigma^\nu
 + A_{11}(q^2)\sigma^\nu \bar{\sigma}^\rho \sigma^\mu 
k_\rho
 + A_{12}(q^2)\sigma^\mu \bar{\sigma}^\rho \sigma^\nu 
q_\rho
 \bigr]_{\alpha \dot{\beta}}\bchF{\dot{\beta}}(\pF), 
\end{array} \label{matrix_Q}
\end{equation}
where $q^\mu = \pB^\mu -\pF^\mu$, $k^\mu = \pB^\mu + 
\pF^\mu$ 
and $\epsilon^{\ast}_\nu(\pB)$ is a polarization vector with 
$\pB \cdot \epsilon^{\ast}(\pB) = 0$. 

Conservation of the supercurrent leads to 
a relation among the form factors 
\begin{equation}
 q^2 \left[ A_{10}(q^2) + A_{11}(q^2)-A_{12}(q^2) \right]
  = -2 \Delta m^2 A_{11}(q^2),\label{current_conservation}
\end{equation}
where $\Delta m^2 \equiv m_{\rm B}^2 - m_{\rm F}^2$ is the 
mass-splitting 
between the gauge boson and the gaugino.

A matrix element of the NG fermion source $j^{\rm 
NG}_{\alpha}(x)$ 
between the gauge boson and the gaugino states are 
decomposed as
\begin{eqnarray}
 \langle \pB |j_{\alpha}^{\rm NG}(0)| \pF \rangle \ = \ 
 \epsilon^{\ast}_\nu(\pB) \left[
   B_{1}(q^{2}) q^{\nu} 
 + B_{2}(q^{2}) q_{\rho}\sigma^{\rho}\bar{\sigma}^{\nu}
 \right]_{\alpha}^{\ \ \beta} \chF{\beta}(\pF)
  \vspace{1mm} \nonumber\\
\hspace{5cm} + \ \epsilon^{\ast}_{\nu}(\pB) \left[
   B_{3}(q^{2})q^{\nu}\sigma^{\rho}q_{\rho} 
 + B_{4}(q^{2})\sigma^{\nu}
 \right]_{\alpha \dot{\beta}}\bchF{\dot{\beta}}(\pF),
\end{eqnarray}
and thus
\begin{eqnarray}
 \langle\pB |\bar{\psi}_{\rm NG}^{\dot{\alpha}}(0)| \pF \rangle 
 \ = \ \epsilon^{\ast}_\nu(\pB) \left[
 - \displaystyle{\frac{B_{1}(q^{2})}{q^{2}}} 
q^{\nu}\bar{\sigma}^{\rho} q_\rho
 + B_{2}(q^{2}) \bar{\sigma}^{\nu}
 \right]^{\dot{\alpha}\beta} \chF{\beta}(\pF)
 \vspace{1mm} \nonumber \\
\hspace{5cm} + \ \epsilon^{\ast}_{\nu}(\pB) \left[
 B_{3}(q^{2})q^{\nu} 
 - \displaystyle{\frac{B_{4}(q^{2})}{q^{2}}}q_{\rho} 
\bar{\sigma}^{\rho}
 \sigma^{\nu}  \right]^{\dot{\alpha}}_{\ \ \dot{\beta}}
 \bchF{\dot{\beta}}(\pF).\label{matrix_lambda}
\end{eqnarray}

The regularity of the form factors of the matrix element for 
$J^\mu_\alpha -\sqrt{2}if \sigma^{\mu}_{\alpha \dot{\alpha}} 
\bar{\psi}_{\rm NG}^{\dot{\alpha}}$ as $q^2 \to 0$ 
leads to the singularity of the form factor $A_{12}(q^2)$ at 
$q^2 = 0$ 
\begin{equation}
 \lim_{q^2 \to 0} q^2A_{12}(q^2) = -\sqrt{2}if B_4(0).
\end{equation}

Substituting it into Eq.(\ref{current_conservation}) 
with the limit $q^2 \to 0 $, we obtain 
\begin{equation}
 \sqrt{2}if B_4(0) = -2 \Delta m^2 A_{11}(0).  \label{B4_A11}
\end{equation}

We can relate 
the form factor $B_4(0)$ to an effective coupling constant of 
 the NG fermion with the gauge boson and the gaugino 
forming 
 a gauge supermultiplet. 
By repeating the same procedure as that in the previous 
subsection 
leading to Eq.(\ref{ME-of-Lint}), we obtain
\be
 \langle \pB|\cL_{\rm int}(0)|q;\pF\rangle_{\rm I}=
 -\chng(q)q_{\mu}\sigma^{\mu}
 \langle \pB|\bpsng(0)|\pF\rangle
 -\bchng(q)q_{\mu}\bar{\sigma}^{\mu}
 \langle \pB|\psng(0)|\pF\rangle. \label{ME-of-Lint2}
\ee

On the other hand, we expect the following nonderivative 
interaction terms 
in the effective Lagrangian\cite{lee-wu}
\begin{equation}
 \cL_{\rm int} = \hef{} \psng \sigma^{\mu\nu}\lm v_{\mu \nu} 
  + {\rm h.c.}+\cdots,
\end{equation}
where $\lm$ is the gaugino field and $v_{\mu \nu}$ is 
the gauge field strength respectively. 
So its matrix element is written as 
\begin{equation}
 \langle \pB|\cL_{\rm int} (0)|q;\pF \rangle_{\rm I} = i\hef{}
  \epsilon^{\ast}_\nu(\pB) p_{{\rm B}\mu} \chi_{\rm NG}(q)
  \sigma^\nu \bar{\sigma}^\mu 
  \chF{}(\pF)
  +i\hef{} \epsilon^{\ast}_\nu(\pB) p_{{\rm B}\mu} 
  \bar{\chi}_{\rm NG}(q)
  \bar{\sigma}^\nu \sigma^\mu 
  \bchF{}(\pF). \label{Lint-gef}
\end{equation}
{}For the case of $m_{\rm F} \neq 0$, 
comparison between Eq.(\ref{ME-of-Lint2}) and 
Eq.(\ref{Lint-gef}) 
after substitution of Eq.(\ref{matrix_lambda}) into 
Eq.(\ref{ME-of-Lint2}) 
gives
\begin{equation}
 B_4(0) = -i m_{\rm F} \hef{}.
\end{equation}
Using Eq.(\ref{B4_A11}) we obtain the analog of the 
Goldberger-Treiman 
relation for gauge supermultiplets
\begin{equation}
 -\sqrt{2}\hef{}f=\frac{2(m_{\rm B}^2 - m_{\rm F}^2)}{m_{\rm F}} 
A_{11}(0). 
 \label{GTR}
\end{equation}

To determine the form factor $A_{11}(0)$, 
we substitute the following expression of the supercurrent 
into Eq.(\ref{matrix_Q}) with the limit $q^2\to 0$. 
\begin{equation}
 J^\mu_\alpha = \sqrt{2}if \sigma^{\mu}_{\alpha \dot{\alpha}}
 \bar{\psi}_{\rm NG}^{\dot{\alpha}} 
 -i 
v_{\nu\rho}\left(\sigma^{\nu\rho}\sigma^{\mu}\right)_{\alpha\dot{\alpha}}
 \bar{\lm}^{\dot{\alpha}} + \cdots \ .
\end{equation}
Then we find 
\begin{equation}
 A_{11}(0)= \frac{1}{2}.
\end{equation}
By substituting it into Eq.(\ref{GTR}), we obtain the 
low-energy theorem for the gauge supermultiplets
\begin{equation}
\label{LETgauge}
 \hef{}=-\frac{1}{\sqrt{2}f} \left( \frac{m_{\rm B}^2}{m_{\rm F}} 
-m_{\rm F} 
 \right).
\end{equation}

\section{Relation among central charge and order 
parameters} \label{V0-f1f2}
In the single-wall case, the order parameter $f$ for the SUSY 
breaking 
due to the existence of a BPS domain wall is given by 
the square root of the energy density of the wall $\sqrt{V_0}$. 
In the two-wall system, however, two different SUSY 
breakings occur, 
whose origins are our wall and the other wall respectively.
Thus there are in general two kinds of order parameters $f_1$ 
and $f_2$ 
for different SUSY breakings. 
Here we shall clarify the relation among $f_1$, $f_2$ and 
$V_0$ and 
the central charge of the SUSY algebra. 

Let us begin with the four-dimensional SUSY algebra of the 
bulk theory.
Since we consider the case of the SUSY breaking, 
we describe the SUSY algebra in the local form.
The three-dimensional SUSY algebra can be derived from 
the four-dimensional one with the central charge:
\bea
 \left\{ Q_{\alpha}, \bar{J}^{\nu}_{\dot{\beta}}(X)\right\}
 &\!\!=&\!\!2\sigma^{\mu}_{\alpha\dot{\beta}}T_{\mu}^{\ 
\nu}(X), 
 \label{Q-Jbar} \\
 \left\{ Q_{\alpha}, J^{\nu}_{\beta}(X)\right\}
 &\!\!=&\!\!4i\left(\sigma^{\mu}\bar{\sigma}^{\nu}\right)_{\alpha}^{\ \gamma}
 \epsilon_{\gamma\beta}\del_{\mu}W^*(A^*(X)), \label{Q-J}
\eea
where $T_{\mu}^{\ \nu}(X)$ is the energy-momentum tensor 
\be
 T_{\mu}^{\ 
\nu}=\del^{\nu}A^*\del_{\mu}A+\del^{\nu}A\del_{\mu}A^*
 +\frac{i}{2}\bar{\Psi}\bar{\sigma}^{\nu}\del_{\mu}\Psi
 +\frac{i}{2}\Psi\sigma^{\nu}\del_{\mu}\bar{\Psi}+\delta_{\mu}^
{\ \nu}\cL
\ee
The term containing the superpotential $W(\Phi)$ represents 
the density of 
the central charge. 
In this section, we calculate the SUSY algebra 
in the following Wess-Zumino model for simplicity.
\be
 \cL=\left.\bar{\Phi}\Phi\right|_{\tht^2\bar{\tht}^2}
 +\left.W(\Phi)\right|_{\tht^2}+{\rm h.c.},
\ee
where $\Phi=A+\sqrt{2}\tht\Psi+\tht^2 F$ is a chiral 
superfield.

Eqs.(\ref{Q-Jbar}) and (\ref{Q-J}) can be rewritten in terms of 
the three-dimensional supercurrent defined by 
Eq.(\ref{3D-supercurrents}) as 
\bea
 \left\{Q^{(1)}_{\alpha},J^{(1)n\beta}(x)\right\}
 &\!\!=&\!\! 2\left(\gm{m}\right)_{\alpha}^{\ \beta}
 \left(Y_{m}^{\ n}-2\delta_{m}^{\ n}\Delta W^* \right), \\
 \left\{Q^{(2)}_{\alpha},J^{(2)n\beta}(x)\right\}
 &\!\!=&\!\! 2\left(\gm{m}\right)_{\alpha}^{\ \beta}
 \left(Y_{m}^{\ n}+2\delta_{m}^{\ n}\Delta W^* \right), \\
 \left\{Q^{(1)}_{\alpha},J^{(2)n}_{\beta}(x)\right\}
 &\!\!=&\!\! 2\epsilon_{\alpha\beta}Y_{2}^{\ n}, \\
 \left\{Q^{(2)}_{\alpha},J^{(1)n}_{\beta}(x)\right\}
 &\!\!=&\!\! -2\epsilon_{\alpha\beta}Y_{2}^{\ n}, 
\eea
where 
\be
 Y_{\mu}^{\ \nu}(x)\equiv \int\dy T_{\mu}^{\ \nu}(X), \;\;\;
 \Delta W\equiv \int\dy \del_{y}W(A(X)).
\ee

Note that $\Delta W$ is constant since it depends only on 
the boundary condition along the extra dimension and 
becomes non-zero on a non-trivial boundary condition.
Here we suppose that the background configuration $\Acl(y)
$ is real, 
for simplicity.
Thus the central charge $\Delta W$ is treated as a real 
constant in the 
following discussion.

Since the background $\Acl(y)$ is real, four-dimensional 
fields $A(X)$ and 
$\Psi(X)$ are mode-expanded as follows.
\bea
 A(X)&\!\!=&\!\!\Acl(y)+\frac{1}{\sqrt{2}}\left(\sum_{p=0}
 ^{\infty}
 \bR{p}(y)\ar{p}(x)
 +i\sum_{p=1}^{\infty}\bI{p}(y)\ai{p}(x)\right), \\
 \Psi(X)&\!\!=&\!\!\frac{1}{\sqrt{2}}\left(\sum_{p=0}^{\infty}
 \f{1}{p}(y)\ps{1}{p}(x)
 +i\sum_{p=0}^{\infty}\f{2}{p}(y)\ps{2}{p}(x)\right). 
\eea
Note the NG boson $\ar{0}(x)$ for the broken translational 
invariance 
along the extra dimension comes from 
 the real part of $A(x)$ because $\Acl(y)$ is real.
In the fermionic sector, there are the NG fermions 
 $\ps{2}{0}(x)$ and $\ps{1}{0}(x)$ for broken $Q^{(1)}$- and 
$Q^{(2)}$-SUSY, respectively.
 
$Y_{\mu}^{\ \nu}(x)$ can be rewritten in terms of 
three-dimensional fields as 
\bea
 Y_{m}^{\ n}(x)&\!\!=&\!\! -\delta_m^{\ n} 
V_{0}+\sum_{p=0}^{\infty}\del^n 
 \ar{p}\del_m \ar{p}
 +\sum_{p=1}^{\infty}\del^n \ai{p}\del_m \ai{p} \nonumber\\
 &\!\!\!&\!\!+\frac{i}{2}\sum_{p=0}^{\infty}\ps{1}{p}\gm{n}\del_m 
\ps{1}{p}
 +\frac{i}{2}\sum_{p=0}^{\infty}\ps{2}{p}\gm{n}\del_m \ps{2}{p}
 +\delta_m^{\ n} \cL^{(3)} \\
 &\!\!=&\!\! -\delta_m^{\ n}V_{0}+T_{(3)m}^{\ \ \ \ n}(x), \\
 Y_2^{\ m}(x)&\!\!=&\!\! f_{\rm P}\del^m \ar{0}+\cdots, 
\label{Y2m}
\eea
where $V_{0}$ is the energy density of the background 
\be
 V_{0}\equiv\int\dy\left\{(\del_{y}\Acl)^2
 +\left|\frac{\del W}{\del A}\right|^2_{A=\Acl}\right\}, 
\ee
and $T_{(3)m}^{\ \ \ \ n}(x)$ is the three-dimensional 
energy-momentum tensor.
$f_{\rm P}$ in Eq.(\ref{Y2m}) corresponds 
an order parameter for the breaking of 
the translational invariance along the extra dimension 
\be
 f_{\rm P}=\sqrt{2}\int\dy \bR{0}(y)\del_y \Acl(y).
\ee

Then, the three-dimensional SUSY algebra becomes as 
follows.
\bea
 \left\{Q^{(1)}_{\alpha}, J^{(1)n\beta}(x)\right\}
 &\!\!=&\!\! 2\left(\gm{m}\right)_{\alpha}^{\ \beta}
 \left\{-\delta_m^{\ n}(V_{0}+2\Delta W)+T_{(3)m}^{\ \ \ \ 
n}\right\}, 
 \label{Q1-Q1_com} \\
 \left\{Q^{(2)}_{\alpha}, J^{(2)n\beta}(x)\right\}
 &\!\!=&\!\! 2\left(\gm{m}\right)_{\alpha}^{\ \beta}
 \left\{-\delta_m^{\ n}(V_{0}-2\Delta W)+T_{(3)m}^{\ \ \ \ 
n}\right\}, 
 \label{Q2-Q2_com} \\
 \left\{Q^{(1)}_{\alpha}, J^{(2)n}_{\beta}(x)\right\}
 &\!\!=&\!\!2\epsilon_{\alpha\beta}\left(f_{\rm P}\del^n \ar{0}
+\cdots \right),
 \label{Q1-Q2_com} \\
 \left\{Q^{(2)}_{\alpha}, J^{(1)n}_{\beta}(x)\right\}
 &\!\!=&\!\!2\epsilon_{\beta\alpha}\left(f_{\rm P}\del^n \ar{0}
+\cdots \right).
 \label{Q2-Q1_com}
\eea

On the other hand, the supercurrents have the following 
forms 
\be
 J^{(1)m}_{\alpha}=\sqrt{2}if_1 
\left(\gm{m}\ps{2}{0}\right)_{\alpha}+\cdots,
 \;\;\;
 J^{(2)m}_{\alpha}=\sqrt{2}if_2 
\left(\gm{m}\ps{1}{0}\right)_{\alpha}+\cdots,
\ee
where $f_1$ and $f_2$ are the order parameters of the 
breaking for 
$Q^{(1)}$- and $Q^{(2)}$-SUSY respectively.

Then using the commutation relation of the 
three-dimensional Majorana spinors 
\be
 \left\{\psi_{\alpha}(\vec{x},t), \psi_{\beta}(\vec{x}',t)\right\}
 =-\left(\gm{0}\sigma^2\right)_{\alpha\beta}\delta^2 
(\vec{x}-\vec{x}'), 
\ee
Eqs.(\ref{Q1-Q1_com}) and (\ref{Q2-Q2_com}) are also 
written as 
\bea
 \left\{ Q^{(1)}_{\alpha}, J^{(1)n\beta}(x)\right\}
 &\!\!=&\!\!-2f_1^2 \left(\gm{n}\right)_{\alpha}^{\ 
\beta}+\cdots, \\
 \left\{ Q^{(2)}_{\alpha}, J^{(2)n\beta}(x)\right\}
 &\!\!=&\!\!-2f_2^2 \left(\gm{n}\right)_{\alpha}^{\ 
\beta}+\cdots.
\eea

By comparing these commutation relations with 
Eqs.(\ref{Q1-Q1_com}) 
and (\ref{Q2-Q2_com}), 
we obtain the following relations 
\be
 V_{0}=\frac{f_1^2+f_2^2}{2},\;\;\; \Delta 
W=\frac{f_1^2-f_2^2}{4}.
 \label{dW_f1-f2}
\ee

Thus 
the average of the squares of 
two different kinds of order parameters gives 
the energy density of the background 
and their difference gives the central charge. 
{}From the second relation in Eq.(\ref{dW_f1-f2}), 
we can conclude that if the extra dimension is 
compactified, the superpotential $W$ must be a multi-valued 
function, 
such as those in Ref.\cite{hofmann}, 
in order to realize a situation where the order parameter for 
the breaking of the $Q^{(1)}$-SUSY is different from 
that of the $Q^{(2)}$-SUSY.

\end{document}